\documentclass[10pt,fleqn,twocolumn]{revtex4}
\usepackage{subfigure,epsfig,graphics,amsmath,rotating}

\begin{document} 

\title{Conductance Properties of Carbon-Based Molecular Junctions} 
\author{Giorgos Fagas$^a$} \thanks{gfagas@nmrc.ie}
\author{Agapi Kambili$^b$} 
\affiliation{
$^a$ NMRC, Lee Maltings, Prospect Row, Cork, Ireland\\
$^b$ Institut f\"{u}r Theoretische Physik, Universit\"{a}t
Regensburg, 93040 Regensburg, Germany}
\date{\today}

\begin{abstract}
We present a comprehensive study of the properties of the
off-resonant conductance spectrum in oligomer nanojunctions between
graphitic electrodes. By employing first-principle-based methods
and the Landauer approach of quantum transport, we identify
how the electronic structure of the molecular junction components
is reflected in electron transport across such systems.
For virtually all energies within the conduction gap of the corresponding
idealised polymer chain, we show that: a) the inverse decay length of
the tunnelling conductance is intrinsically defined by the complex-band
structure of the molecular wire despite ultrashort oligomer lengths of few
monomer units, and b) the contact conductance crucially depends on
both the local density of states on the metal side and the realised
interfacial contact.
\end{abstract}
%\pacs{PACS:}
\maketitle 
\pagestyle{empty}

\section{Introduction}
\label{sec:S1}

Molecular electronics has been an active research field
\cite{Nat00JGA,Sci03NR,JCM03S} since the early seventies,
when a rectifier based on a single organic molecule was
proposed \cite{CPL74AR}. Extensive experimental
\cite{Sci97RZM,JCP98TDH,Sci01CPZ,Nat02SNU,APL03RBW,Sci03KDH,NL03KNY,
PPC03PRL, APL01SHI,JPCB02WHR,JPCB02CPZ,JPCB02AM,JPCB02IMA,PRB03WLR,Sci03XB}
and theoretical \cite{PRB96STD,PRL00VPL,PRB01FCR,PRB601LA,
PRB01KB,JCP02NGI,JCP02LWF,CP02CFR, EPL03GFR,PRB03CLW,PRB03XR,PRL03EK,JPC03MND,
PRL03HWW,JPCB03TY,CMS03STB,CMS03NLG,PRB03KLG,JCP04TS,JACS02SDB,CPC03GGS}
investigations have examined the electronic response of a
single molecule hybridising with metals which act as
donor and acceptor reservoirs.
Electron transport across metal/molecule/metal junctions (MMMs)
depends crucially on the details of the contacts and
the exact choice of materials, so that the fundamental mechanisms
have been highly controversial. Fine tuning may lead to a range of
behaviours in the current-voltage characteristics \cite{Sci03NR}.
Examples include negative differential resistance, strong rectification,
molecular memory phenomena, and even the Coulomb blockade effect,
a hallmark of mesoscopic physics.

For a class of MMMs increasing consensus based on experimental
findings \cite{APL01SHI,JPCB02WHR,JPCB02CPZ,JPCB02AM,JPCB02IMA,PRB03WLR,Sci03XB}
points to a dominant transport mechanism, namely, to
through-bond tunnelling. Such setups typically consist of relatively short oligomers like alkane \cite{APL01SHI,JPCB02CPZ,PRB03WLR,Sci03XB}
and phenyl-based chains \cite{JPCB02WHR,JPCB02AM,JPCB02IMA}
which are covalently bonded either directly or via anchor
groups to the metallic electrodes. The metal of choice is usually Au
for its noble properties. Most recently, junctions with graphitic
and mercury drop contacts have also been investigated \cite{JPCB02AM}.

Experimental results show temperature independent
electron transport and a conductance $\mathrm g$
which decreases exponentially versus the molecular length L,
with an inverse decay length $\beta$ deriving from the oligomer.
Namely,
\begin{equation}
{\mathrm g(E_F,V)}={\mathrm g}_o(E_F,V) {\mathnormal e^{-\beta(E_F,V) L}},
\label{eq1}
\end{equation}
where $E_F$ denotes the equilibrium Fermi energy. The pre-exponential
factor ${\mathrm g}_o$ is determined by the actual interfacial contact.
In most realisations, measured quantities do not depend on the
source-drain voltage $V$ for low bias. In another experiment \cite{PRB03WLR},
a weak dependence has been indicated. Nevertheless, such behaviour
is consistent with the picture of coherent off-resonant
tunnelling across the molecular bridge.

Eq. \ref{eq1} has been theoretically investigated
deep inside the tunnelling regime, both analytically
for $\pi$-electron models of increasing complexity
\cite{JCP94MKR,SSC98OKM,PRB98MJ}
and numerically for some realistic MMMs \cite{PRB03KLG,EPL96JV,PRB97MJ}.
In those studies, electron-vibrational coupling has been disregarded
and the zero temperature $T$, zero bias limit have been taken.
Up to $T$ not much lower than room temperature ($\leq 25$meV),
all assumptions constitute a reasonable approximation for
specific molecular junctions. This depends on the molecular
vibrational frequencies and conformational barriers,
electron resident times, and the non-equilibrium electrostatic potential profile.
Moreover, these assumptions are in agreement with the above experiments,
allowing for an electrical response that is predominantly a property of
equilibrium conditions, such as the band line-up or
charge relaxation
\cite{PRB601LA,JCP02NGI,JCP02LWF,PRB03CLW,PRB03XR,JPC03MND,CMS03STB},
the optimised atomic configuration of the molecular bridge
\cite{PRB96STD,CMS03NLG,PRB03KLG,JCP04TS}, and the contact microstructure
\cite{PRL00VPL,PRB01KB,PRB01FCR,CP02CFR,EPL03GFR,JACS02SDB,
PRL03EK,PRL03HWW,JPCB03TY,CPC03GGS}.

Electron transport across a molecular junction has a lot in common
with the intramolecular or donor-bridge-acceptor (DA)
electron transfer. The exponential dependence of the conductance
in Eq. \ref{eq1} is reminiscent of the well-studied McConnel's superexchange
mechanism in the weak coupling limit \cite{JCP61McC,JCP92EK,JPPA94RH}, where
\begin{equation}
{\mathrm k_{DA}}=\frac{2\pi}{\hbar} \left| {\mathrm  H_{DA}} \right|^2 F \sim
e^{-\beta(E)L}.
\label{eq1a}
\end{equation}
A connection between the electron transfer rate $k_{DA}$ and the conductance
has been given by Nitzan \cite{JPC01N} in certain limits.
${\mathrm H_{DA}}$ is the coupling between
donor and acceptor electronic states at a potential surface
crossing energy $E$, and $F$ is the Franck-Condon factor.
However, the continuum of states
of the reservoirs and their constraints redefine the problem and have to be
explicitly considered for a full account of the conduction
properties. From this viewpoint, a transport
calculation across the whole system is required.

Mujica {\it et al} \cite{JCP94MKR} generalised previous analytical results of
Everson and Karplus \cite{JCP92EK} on electron transfer across a single-orbital
tight-binding linear homogeneous chain, to study conductance properties.
Their studies revealed the exponential length
decrease for energies far-off the energy spectrum of the molecular
bridge. Onipko {\it et al} \cite{SSC98OKM} found an approximate solution for
conjugated oligomers in the $\pi$-electron model. By disregarding
the effect of the electrodes
to the molecular electronic structure in the tunnel limit
$\beta(E)L \gg 1$, their results closely resemble the
electron transfer problem. However, they provide a simple expression
for ${\mathrm g}_o$. Using extended H\"{u}ckel calculations,
Joachim with Vinuesa \cite{EPL96JV} and Magoga \cite{PRB97MJ} have studied
numerically the conductance at $E_F$ of several relatively long conjugated
molecules, where $E_F$ falls in the middle of the wire conduction gap.
They established that the damping factor is a characteristic
property of the wire and that ${\mathrm g}_o$ is related to the
contact realisation of the bridge, in agreement with Onipko
{\it et al} \cite{SSC98OKM}. Moreover, an analytic expression for $\beta$ was
derived within a simple model \cite{PRB98MJ} and its properties were studied
within a level-repulsion approach of Random Matrix Theory
\cite{CP98GPB,CPL03LJ}.

Most recently, Joachim and Magoga \cite{CP02JM},
and Tomfohr and Sankey \cite{PRB02TS},
provided independently approximate expressions for the effective mass
of electrons in MMMs tunnelling through the oligomer-induced barrier.
They considered the ideal polymer limit and approximated
$\beta$ to the spatial decay parameter of the wavefunction.
This possibility has been conjectured in many of the aforementioned studies.
Tomfohr and Sankey studied the local DOS decay parameter of octanedithiols
between Au (111) surfaces and found good qualitative agreement
with the imaginary part of the wavevector of wavefunctions in
the forbidden energy domain of the corresponding one-dimensional crystal,
namely, the complex-band structure \cite{PR59K,PPS63H}.

In this paper, we examine off-resonant electron transport across
molecular junctions in view of the new experiments and advances
of theoretical concepts. In particular, we provide a comprehensive
analysis of the entire conductance spectrum inside the
Highest-Occupied-Molecular-Orbital--Lowest-Unoccupied-Molecular-Orbital
(HOMO-LUMO) gap $E_g^M$ of phenyl-based oligomers,
as a function of the number of monomer units.
We focus on studies of oligo-phenyl-ethynyl (OPE), planar and non-planar
oligo-para-phenyl (OPP) molecular wires covalently
bonded to electrodes consisting of graphitic ribbons.
First-principle-based methods
are employed to calculate the electronic structure of the molecular
junction components, and the conductance in the Landauer
approach of quantum transport.

We find that the conductance is almost always an exponentially
decreasing function of length and
{\it not only for $\beta(E)L \gg 1$}.
We observe tunnelling characteristics for ultrashort oligomer
molecular junctions and/or energies not far-off the edges of
the conduction gap $E_g$ of the corresponding polymer
(PPE and PPP, respectively).
The contact conductance can be approximately written as
${\mathrm g}_0(E)=\Gamma^2_L\Gamma^2_Uf(E)$, where $\Gamma_{L/U}$
describes the atomic coupling at the molecule/metal interface,
and $f(E)$ reflects both the electrode and molecular spectral properties.
The inverse decay length is determined by the energy spectrum of the
polymer when extended to include complex wavevectors. A remarkable
one-to-one correspondence to the imaginary part of the latter is
found for all cases. A brief account of the complex wavevector
mapping applied to PPE has been previously given by the authors \cite{CPL04FKE}.
Mathematically, the problem of scattering through finite
periodic systems has been studied to some extent in Ref. \cite{JPA99BG}.

Since interfacial barriers in our systems are generally lower than
the ones realised in the experiments, these results explain qualitatively
the observations in molecular junctions of a handful of monomers.
In addition, they acquire special importance when recognising
that $E_F$ may lie anywhere inside the gap.
It is in fact common knowledge that it is difficult to {\it a priori}
locate $E_F$ which crucially depends on the combination of anchor
groups and electrodes.

The choice of OPP and OPE as molecular bridges in our MMMs is
dictated by the usual experimental setups.
Graphene-like macromolecules have been synthesised \cite{CEJ02SBB}
and studied theoretically \cite{JPCB03TY}
as the active components of molecular junctions.
Moreover, driven by a vision of purely carbon electronics experiments with
pyrolitic graphite as one of the metals have been reported
\cite{JPCB02AM}, for which the structures we study
constitute a first approximation.
On the other hand, the graphitic ribbons we consider
have some special electronic states
\cite{JPSJJ96FWN,PRB96NFD,PRB99MNF,JPSJ01SLC,PRB00KMS} with large local
density of states around $E_F$, which are responsible for
additional features in the conductance. In analogy to other
carbon-based systems \cite{EPL03GFR} in which they give much richer spectra,
these states provide an extra channel for resonant transport within
the HOMO-LUMO gap of the molecular bridge \cite{MS04FGR}.
As a function of the length of oligomers,
they also compete against the exponential law Eq. \ref{eq1}.

The structure of the paper is as follows. In
Sec. \ref{sec:S2} we briefly introduce the theoretical methods that
we used, and the details of the investigated systems.
In Sec. \ref{sec:S3} we present the electronic structure
of the studied molecular junctions with emphasis on the properties
relevant to the aimed transport. We also analyse the concept of
complex-band structure. The conductance properties are discussed
in Sec. \ref{sec:S4}. Finally, we make some concluding remarks.

\section{Computational Framework}
\label{sec:S2}

\subsection{Electronic Structure and Transport Properties}
\label{sec:S2A}

The electronic structure of the oligomer as well as that of the electrodes
is treated within an approach based on Density Functional Theory (DFT)
in the local density approximation (LDA).
By employing a linear combination of atomic orbitals,
the method falls into the class of tight-binding (TB)-DFT.
It has been successfully applied to the calculation of
properties of a range of materials including semiconductors,
carbon nanotubes, fullerenes, DNA and proteins
\cite{TBDFT,PRB98EPJ,EPL03GFR,SM03PMC,JMS04EFS}.

The single-particle electronic Kohn-Sham eigenstates $\psi_i$ of the system
are expanded in a non-orthogonal basis set
$\varphi_\mu({\mathbf r-R_\mu})$ taken as a
valence basis localised at the ionic positions ${\mathbf R_\mu}$, namely
\begin{equation}
\psi_i({\mathbf r})=\sum\limits_\mu c^i_\mu\varphi_\mu({\mathbf r-R_\mu}).
\label{eq2}
\end{equation}

When this {\it Ansatz} is substituted in the Kohn-Sham
equations for $\psi_i$, it yields a set of algebraic equations
\begin{equation}
\sum\limits_\nu(H_{\mu\nu}^{TB}-S_{\mu\nu}^{TB}E_i)c^i_\nu=0.
\label{eq3}
\end{equation}

The crucial step in deriving Eq. \ref{eq3} is a further approximation of
the full many-body problem by a tight-binding
Hamiltonian which is a function of atomic densities \cite{PRB98EPJ}.
The {\it a priori} parametrisation scheme for the two-centre,
distance dependent Hamiltonian, $H_{\mu\nu}^{TB}$, and overlap, $S_{\mu\nu}^{TB}$,
matrix elements provides an efficient algorithm for electronic
structure calculations. Explicit computational details
are given in Ref. \cite{TBDFT}.
Even though the form of Eq. \ref{eq3} is evidently that of
the extended H\"{u}ckel, all necessary matrix elements are determined
without introducing empirical parameters.

To calculate the conductance, we adopt the Landauer picture of
quantum transport \cite{Datta,ARPC01N} in combination with the TB-DFT.
In this formalism, the current through a structure
is expressed in terms of the transmission function
$\bar{\mathrm T}(E,V)$.
The latter denotes the probability of electrons with energy $E$
to propagate coherently across the system between the two electron
reservoirs. Reading explicitly,
\begin{equation}
I=\frac{2e}{h}\int_{-\infty}^{+\infty}dE \;\; \bar{\mathrm T}
(E,V)[f(E-\mu _1,T) - f(E-\mu _2,T)],
\label{eq4}
\end{equation}
where $f(E-\mu,T)$ is the Fermi function, and $\mu_{1,2}$ refer to the
chemical potentials of the two electrodes (1,2). The factor $2$
in the equality implies spin degeneracy.

Here, we take the low bias limit and drop the voltage dependence
of $\bar{\mathrm T}$, which in general is a function of the
exact electrostatic profile. This approximation yields
accurate results for $V\sim0.1$eV \cite{PRB03XR}. It is also expected
to hold for a wider bias range for voltages dropping mostly at the contacts.
Finally, the transmission function is given by the Green's function method
of quantum transport via
\begin{equation}
\bar{\mathrm T}(E_F)=
Tr \left[ {\mathbf \Gamma}_1{\mathbf G}_M^r{\mathbf \Gamma}_2 {{\mathbf G}_M^r}^{\dag}  \right],
\label{eq5}
\end{equation}
where ${\mathbf G}_M ^{r}(E)=
(E \mathbf S_M^{TB}-\mathbf H_M^{TB}- {\mathbf \Sigma)^{-1}}$
is the retarded molecular Green function at equilibrium.

Using the L{{\"o}}wdin projection technique,
${\mathbf G}_M ^{r}$ is dressed by a self-energy interaction
${\mathbf \Sigma} = {\mathbf \Sigma_1} + {\mathbf \Sigma_2}$ that
takes into account the hybridisation of the molecular
bridge with the metals. The spectral widths
${\mathbf \Gamma}_{1,2}$ act as an effective
coupling to the electronic states of the electrodes.
${\mathbf \Sigma}_{1,2}$ and ${\mathbf \Gamma}_{1,2}$ are given by
\begin{equation}
{\mathbf \Sigma}_{1,2}(E) = {\mathbf J}_{1,2-M}^{\dag}(E)
\:{\mathbf G}_{1,2}^r(E) \:{\mathbf J}_{1,2-M}(E)
\label{eq6}
\end{equation}
and
\begin{equation}
{\mathbf \Gamma}_{1,2} (E)= i \left[ {\mathbf \Sigma}_{1,2}(E) - {\mathbf \Sigma}_{1,2}^{\dag} (E) \right],
\label{eq7}
\end{equation}
respectively.
${\mathbf J}_{1,2-M}=E\mathbf S_{1,2-M}-\mathbf H_{1,2-M}^{TB}$ is
the generalised expression for a non-orthogonal basis,
instead of the simple occurrence of the
coupling matrix $\mathbf H_{1,2-M}^{TB}$
between the electrodes and the molecule.
The retarded Green function ${\mathbf G}_{1,2}^r$ of
the leads is calculated by the Sanvito {\it et al} algorithm \cite{PRB99SLJ},
also extended to include the overlap between atomic orbitals.
The above procedure mainly involves the manipulation of
$N_M\times N_M$ matrices, where $N_M$ is the number of atomic
orbitals taken at positions on the molecule. Exception
is Eq. \ref{eq6} that includes electrode degrees of freedom.
However, interactions are of short-range due to the
tight-binding approximation, and matrix
dimensions never exceed $\sim N_M$.

In what follows, we focus on generic properties
of electron transport across carbon-based molecular junctions.
To this end, we study the conductance spectrum which is defined via
\begin{equation}
{\mathrm g}(E)=\frac{2e^2}{h}\bar{\mathrm T}(E).
\label{eq8}
\end{equation}
It follows from Eq. \ref{eq4} within the linear response
that the above expression at $E_F$ strictly coincides with the
conductance for very low bias. For finite but small voltages
such that non-equilibrium effects are insignificant
generalised formulae involving ${\mathrm g}(E)$
may be derived (cf \cite{JCP98TDH}).
However, a complete investigation of the voltage dependence
is out of the scope of the paper and is left for the future.
We note the main effect of moderate-to-room temperature
is to broaden features of the conductance at the $\sim0.1$eV scale.

\begin{figure}[h]
\centerline{\epsfig{figure=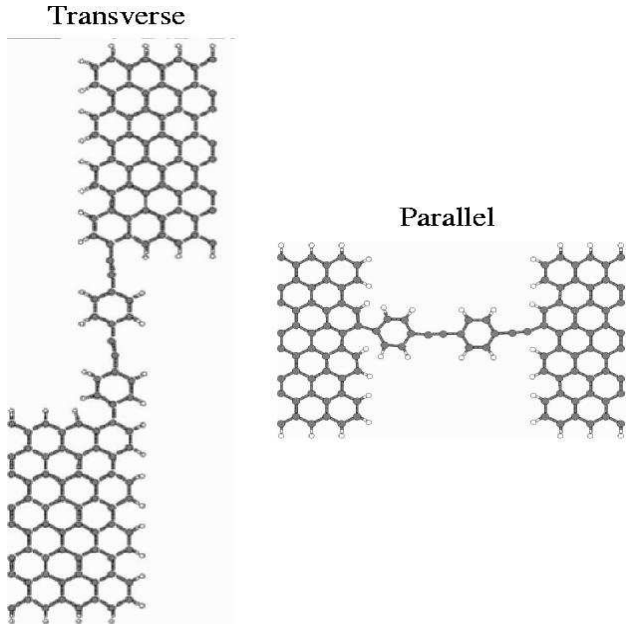,
width=.95 \linewidth ,height=.95 \linewidth}}
\caption{A representative molecular junction among those studied
is the structure of diphenyl-ethynylene
($N=2$) bonded to two graphitic electrodes along either the zigzag (left)
or the armchair (right) edge. Open circles denote hydrogen atoms.
See Sec. \ref{sec:S2B} for more details.}
\label{f1}
\end{figure}

\subsection{System Specifics}
\label{sec:S2B}

We consider all-carbon molecular junctions such
as those depicted in Fig. \ref{f1}.
The electrodes are (zigzag) graphitic ribbons of the indicated
width and infinite extension along the carbon boundary of zigzag shape.
By taking into account the weak inter-plane interactions in
graphite, we isolate a single plane for simplicity.
The molecule is a conjugated molecular wire of a few
repeated monomer units, $N$, also taken as our length unit.
To demonstrate topological effects on
the conductance, we initially examine two different types
of junctions; one with the oligomer being attached
to the electrodes along the zigzag edge
('transverse' configuration), and one with the wire
connected to the armchair side of the ribbons ('parallel' configuration),
as shown in Fig. \ref{f1}. All dangling bonds at edge sites are saturated
by hydrogen atoms (open circles in Fig. \ref{f1}), so that they do not
contribute to the density of states around the Fermi energy $E_F$.

\begin{figure}[h]
\centerline{\epsfig{figure=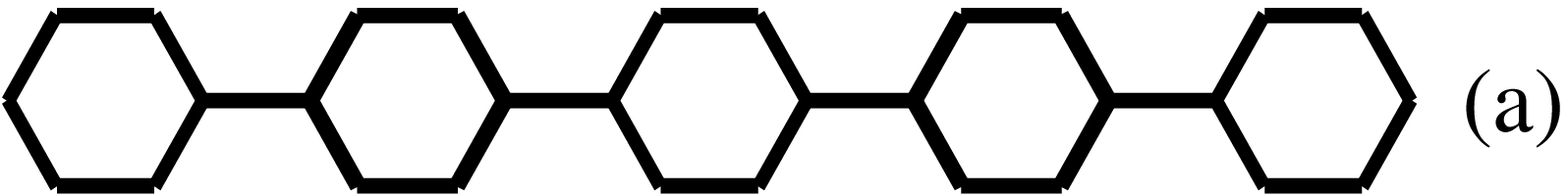, width=.9\linewidth}}
\vspace{0.5cm}
\centerline{\epsfig{figure=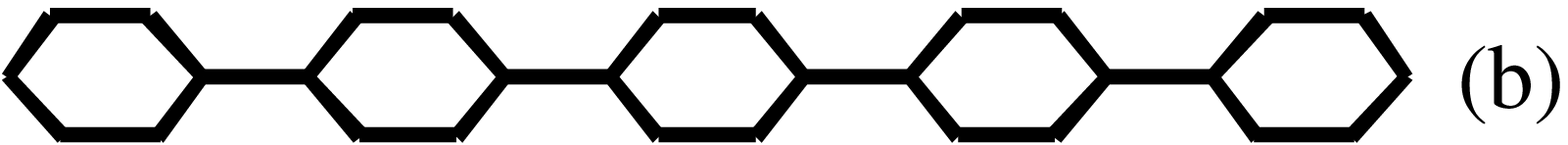, width=.9\linewidth}}
\vspace{0.5cm}
\centerline{\epsfig{figure=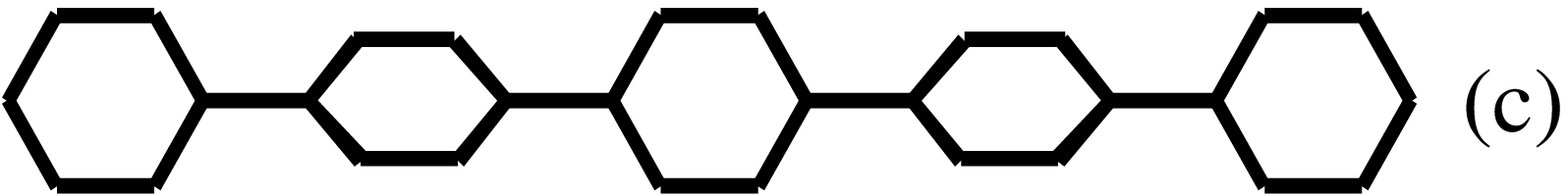, width=.9\linewidth}}
\vspace{0.5cm}
\centerline{\epsfig{figure=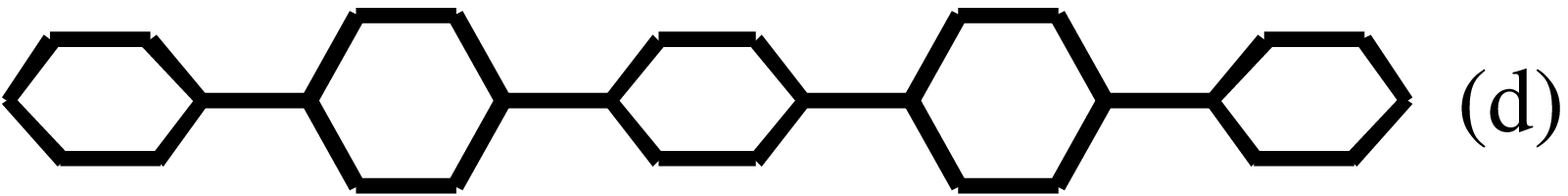, width=.9\linewidth}}
\vspace{0.5cm}
\centerline{\epsfig{figure=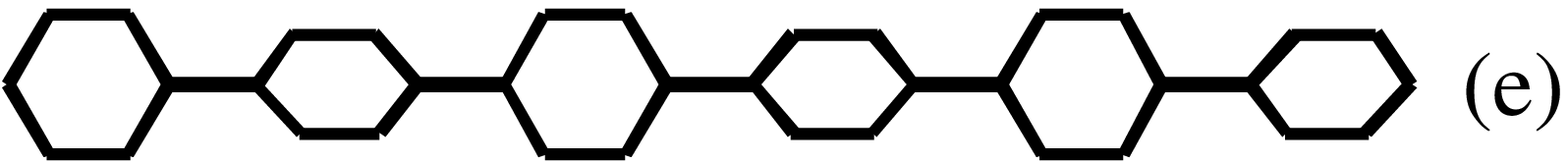, width=.9\linewidth}}
\caption{Configurations of oligo-para-phenyl molecular
bridges studied in this paper. They are distinguished between
odd and even $N$, e.g. for (a)-(d), $N=5$, and
for (e), $N=6$, respectively. The graphene ribbon electrodes
contacted at the end-sites on both sides are in-plane with the page.
The tetrahedral angle for non-planar OPP ((c)-(e)) and the contact geometry (b)
is $45^{\circ}$. }
\label{f2}
\end{figure}

The molecular bridges under investigation are formed by two
chemically different oligomers, and conformationally varying:
a) oligo-phenyl-ethynylene, and b) planar and non-planar
oligo-para-phenylene which differs from OPE in that it
does not have the triple carbon-carbon bond between the benzene rings.
This makes OPP more flexible with
energetically favourable conformation the non-planar one.
Tetrahedral angles between benzene rings may vary
from $45^{\circ}$ for very small oligomers to $27^{\circ}$
for polymers \cite{PRB95AMV,JCPA01GGR}. At room temperature, ring
rotational motion occurs between the planar and
non-planar conformation. When the temperature is cooled down
both possibilities may be realised in experiments with
self-assembled monolayers \cite{JPCB02AM}. Therefore, we consider several
geometries as summarised in Fig. \ref{f2}.

The structures (molecular bridge+electrodes)
are optimised in all calculations. To simulate the bulk
electrodes, a constrained atomic optimisation of junctions
such as those of Fig. \ref{f1} is performed.
Graphene atoms away from the oligomer/ribbon
interfacial bond by approximately $3 \times$hexagonal lattice constant
remain immovable in the positions they assume in perfect graphene.
The inner carbon atoms, which look unsaturated in Fig. \ref{f1},
are bonded to H to avoid spurious effects during optimisation.
As input, initial single C-C bond lengths equal $1.39$\r{A},
triple C-C bond lengths equal $1.2$\r{A},
and C-H distances equal to $1.09$\r{A}. All angles are at $120^\circ$.

\section{Electronic structure}
\label{sec:S3}

\subsection{Electrodes: Graphene Ribbons}
\label{sec:S3A}
There has been an extensive interest in the properties
of nano-graphitic materials recently.
Specifically, ribbon-shaped graphite as in Fig. \ref{f1}
is intensively investigated and the scope of this
section is limited to pointing out the electronic
properties that are crucial for the conductance of
the systems we study. These are related to peaks
in the density of states (DOS).
Such features arise from the low dimensionality
of the ribbons, the topology of their edges,
and the hexagonal lattice.

\begin{figure}[h]
\centerline{\epsfig{figure=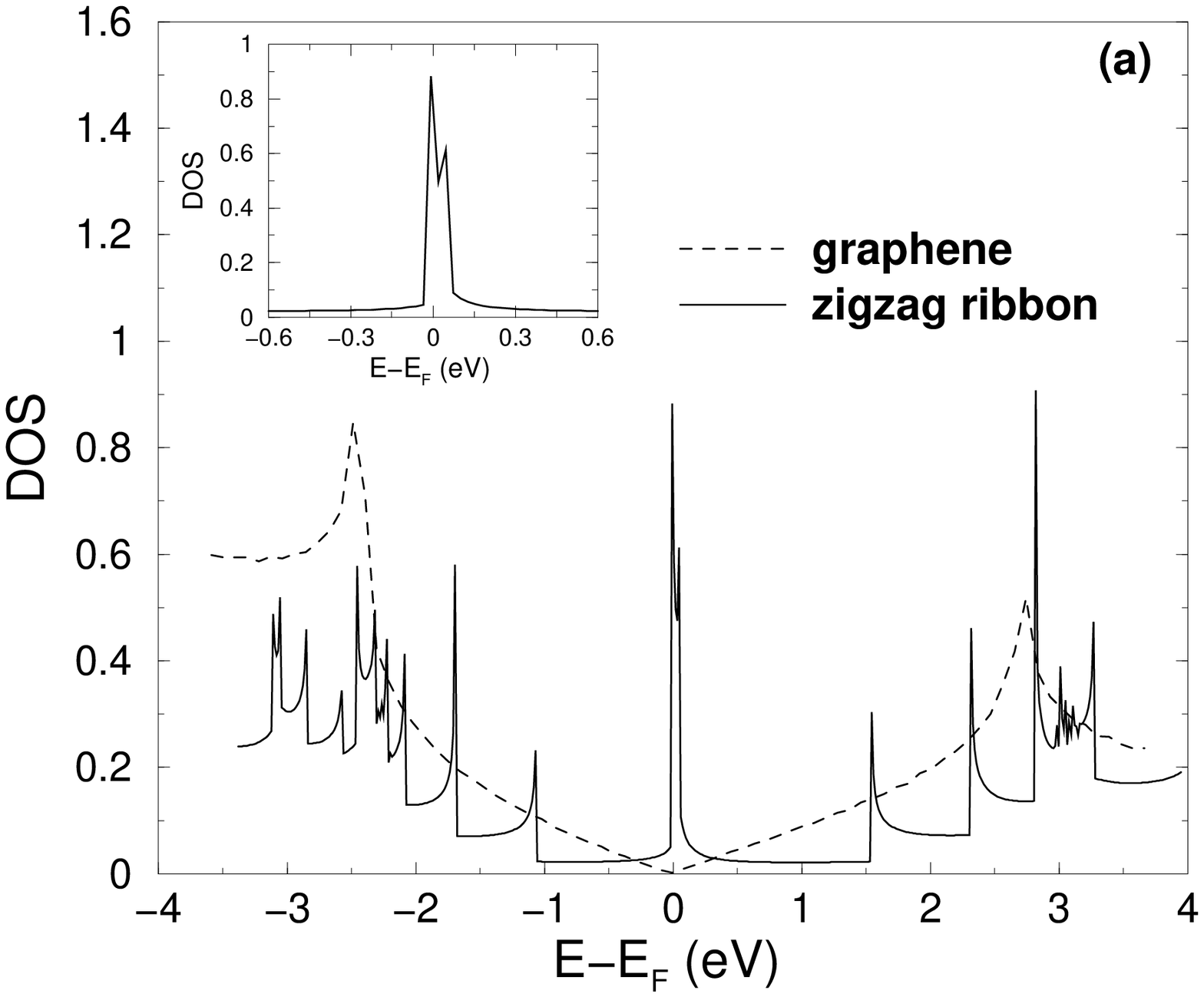,width=14pc,height=14pc}}
\centerline{\epsfig{figure=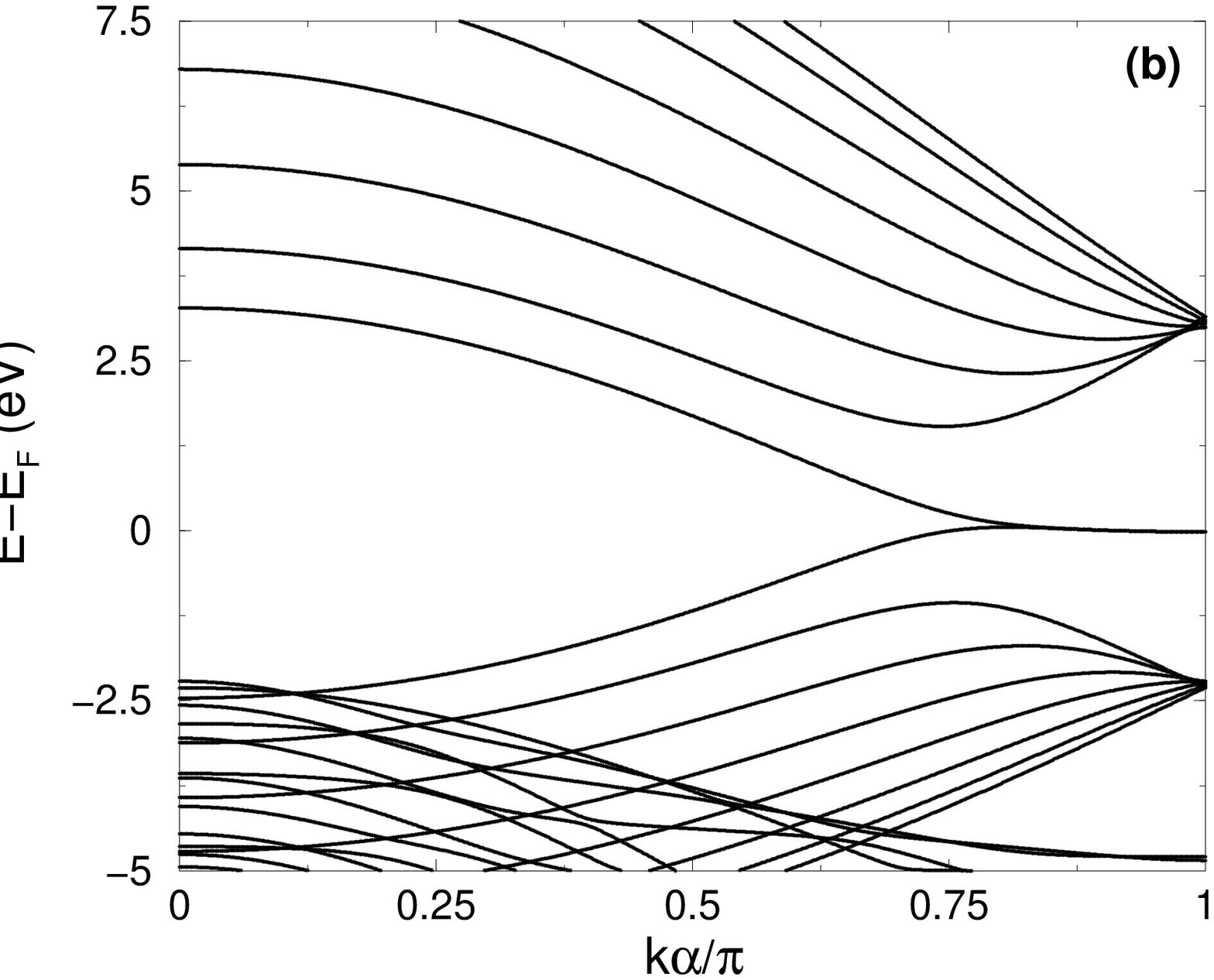,width=14pc,height=14pc}}
\caption{(a) Density of states (DOS) for graphitic systems;
dashed and solid lines correspond to graphene and a zigzag
ribbon, respectively.
The inset shows the DOS of the ribbon at energies around
$E_F$. (b) Band structure of a zigzag ribbon of width $M=4$.}
\label{f3}
\end{figure}

In Fig. \ref{f3}(a) we plot the density of states
of a ribbon with width $M=4$. The latter is
given by the maximum number of honeycomb cells that can be found
in a cross-section.
The ribbon DOS is contrasted with the DOS of bulk graphene (dotted line).
It exhibits the characteristic van-Hove singularities
of quasi-one-dimensional systems. Apart from that one
around $E_F$, which we discuss separately, their
peak positions depend on the width, and they
are present independent of the topology of the edges.

The double peak of the ribbon DOS around $E_F$ (inset
of Fig. \ref{f1}) is generic to graphene edges with a zigzag
geometry \cite{JPSJJ96FWN,PRB96NFD}.
For zigzag ribbons, which are periodically repeated
structures, this develops from two partially flat bands
at the Fermi level, conferring metallic properties \cite{JPSJ01SLC}.
To rationalise the above features, the band structure is
shown in Fig. \ref{f3}(b), which is in good agreement with
less approximate plane-wave DFT calculations \cite{PRB99MNF,PRB00KMS}.

The electronic states corresponding to the low energy bands
are termed edge states. This reflects the property of
wavefunction localisation near the edges.
In fact, in a seminal paper of Lee and Joannopoulos \cite{PRB81LJ} on the
calculation of surface states, the honeycomb lattice with
zigzag edges was presented
as a paradigm. Other specific studies on tight-binding models
\cite{JPSJJ96FWN,PRB99MNF,JPSJ01SLC} of graphene revealed a
single peak arising at $E_F$,
which owes to the simplicity of the $\pi$-electron Hamiltonian.
In those, the topological origin of the peak and its
stability against disorder were also established.

Finally, we note that for applications in molecular electronics,
one should, more precisely, look at the local DOS since the local
electronic structure is most important. The overall DOS may be misleading.
For example, electronic states in a quasi-one-dimensional
square lattice or simply those of the Anderson-Newns model of chemisorption
possess van-Hove singularities but exhibit a smooth
local DOS. This owes to a cancellation of border zone anomalies
(group velocity $v_g \propto sin(k\alpha)$) by the longitudinal
component of the wavefunction \cite{CP02CFR}.
Nevertheless, this does not apply to our discussion.
Using the analogy of graphene ribbons to
carbon nanotubes, and in agreement with the topological arguments,
we employ a previous result \cite{CP02CFR} stating that for states with
zero group velocity outside the border zone, as in Fig. \ref{f3}, no
such cancellation occurs. Indeed, we have
calculated the local density of states at the carbon atoms on
the electrode sides which are bonded to the oligomers
for all studied molecular junctions. Unless stated, we have found small
deviations from the DOS of Fig. \ref{f3} (see also inset of Fig. \ref{f5}),
mainly caused by the oligomer chemisorption.

\subsection{Molecular Wires: Poly-Phenyl-Ethynylene (PPE)
and Poly-Para-Phenylene (PPP)}
\label{sec:S3B}

A core result of this paper relates to the
fingerprints of the electronic band structure of the perfect
one-dimensional crystals that correspond to
the molecular bridges of Fig. \ref{f1} and \ref{f2}, namely,
PPE and PPP, respectively.
In particular, since we study off-resonant electron transport within the
HOMO-LUMO gap, we are not interested in the usual band structure defined
for real Bloch vectors $k$. As explained below,
we need instead to quantify the spatial decay of electronic
evanescent states for energies within the energy gap
$E_g$ between the top of the valence, $E_v$, and the
bottom of the conduction, $E_c$, bands. The latter develop
from the HOMO and LUMO, respectively, of long wires.

When  molecular junctions are formed the molecular electronic states
hybridise with the metal wavefunctions giving a finite
width in energy space and a renormalised HOMO-LUMO gap $E_g^M$ ($\geq E_g$).
For energies within the gap, matching to the metal
side implies decaying wavefunctions within the molecule.
These states are the analogue of the metal-induced gap states (MIGS)
in metal-insulator (or metal-semiconductor) junctions \cite{PPS63H}.
MIGS are responsible for the overall finite DOS
within the $E_g^M$ for short molecular wires because of
contributions coming from both electrodes \cite{PRB601LA}.
For longer wires, the MIGS are mainly located near the
interfaces on either side of the molecule.
Indeed, by looking at the positional dependence of the local
DOS, such behaviour has been observed for
Si nanowires between Al electrodes \cite{PRL00LBS} and alkanethiol chains
between Au (111) surfaces \cite{PRB02TS,JCP03PSS}. We come back to these
points in Sec. \ref{sec:S4B}.

By assuming that interfacial effects at a molecular junction are not dominant,
we turn our
attention to decaying states in the idealised underlying periodic systems
of PPE and PPP. In this case, it is well known that
wavefunctions vanish exponentially in space in the forbidden energy domain.
The decay parameter is quantified by the imaginary component
of Bloch vectors ($k \equiv  q - \imath \kappa$). The latter are obtained
via the method of the complex-band structure, which may be understood as
an extension of the conventional band structure to the complex plane
\cite{PR59K,PPS63H}. Computational schemes are intimately related
to the quest for surface states \cite{PRB81LJ} and the method is as follows.

\begin{figure}[h]
\begin{minipage}{0.5pc}
\begin{rotate}{90}
E-E${\rm_F}$
\end{rotate}
\end{minipage}
\begin{minipage}{15pc}
\centerline{\epsfig{figure=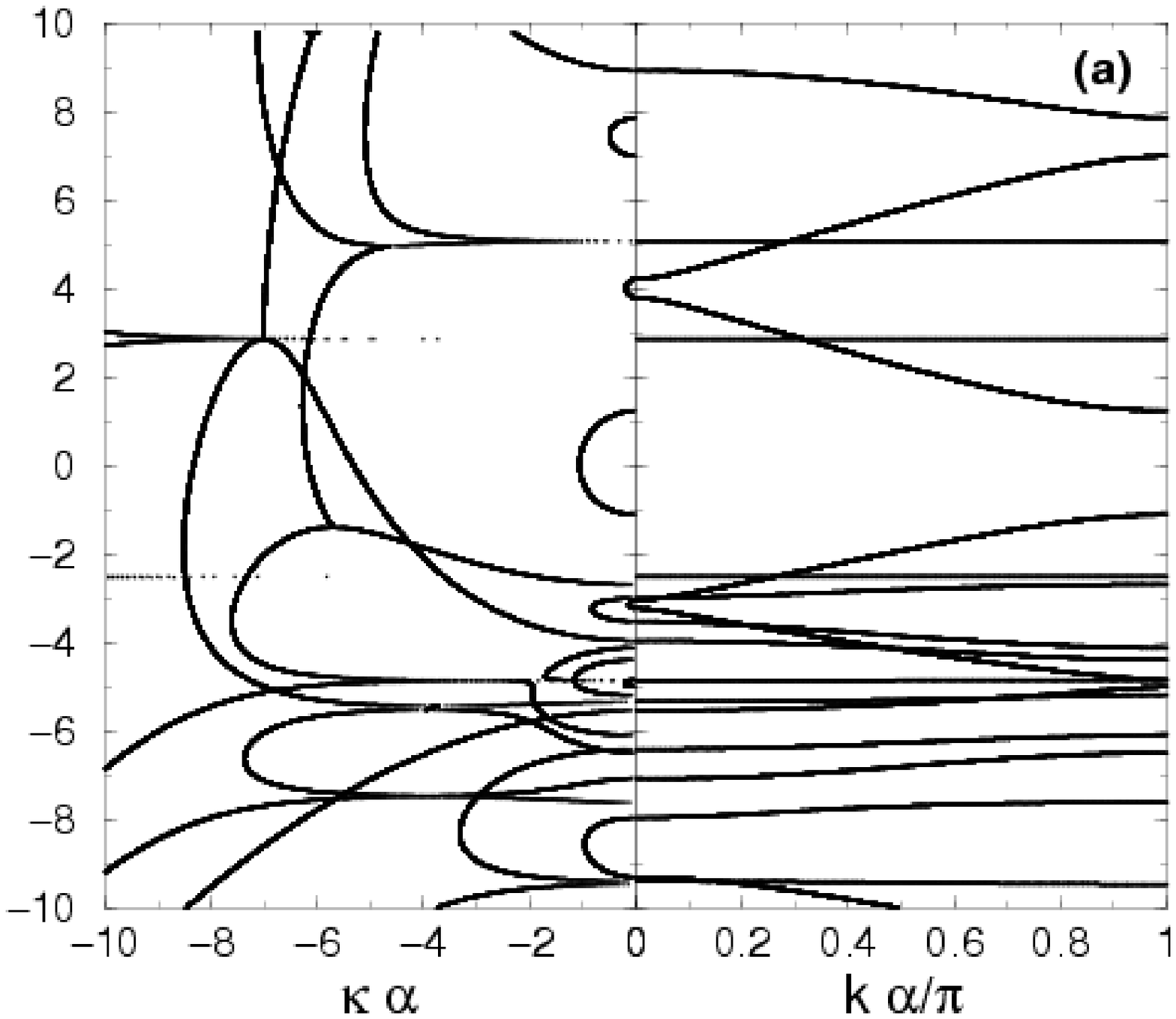,width=14pc,height=14pc}}
\end{minipage}

\begin{minipage}{0.5pc}
\begin{rotate}{90}
E-E${\rm_F}$
\end{rotate}
\end{minipage}
\begin{minipage}{15pc}
\centerline{\epsfig{figure=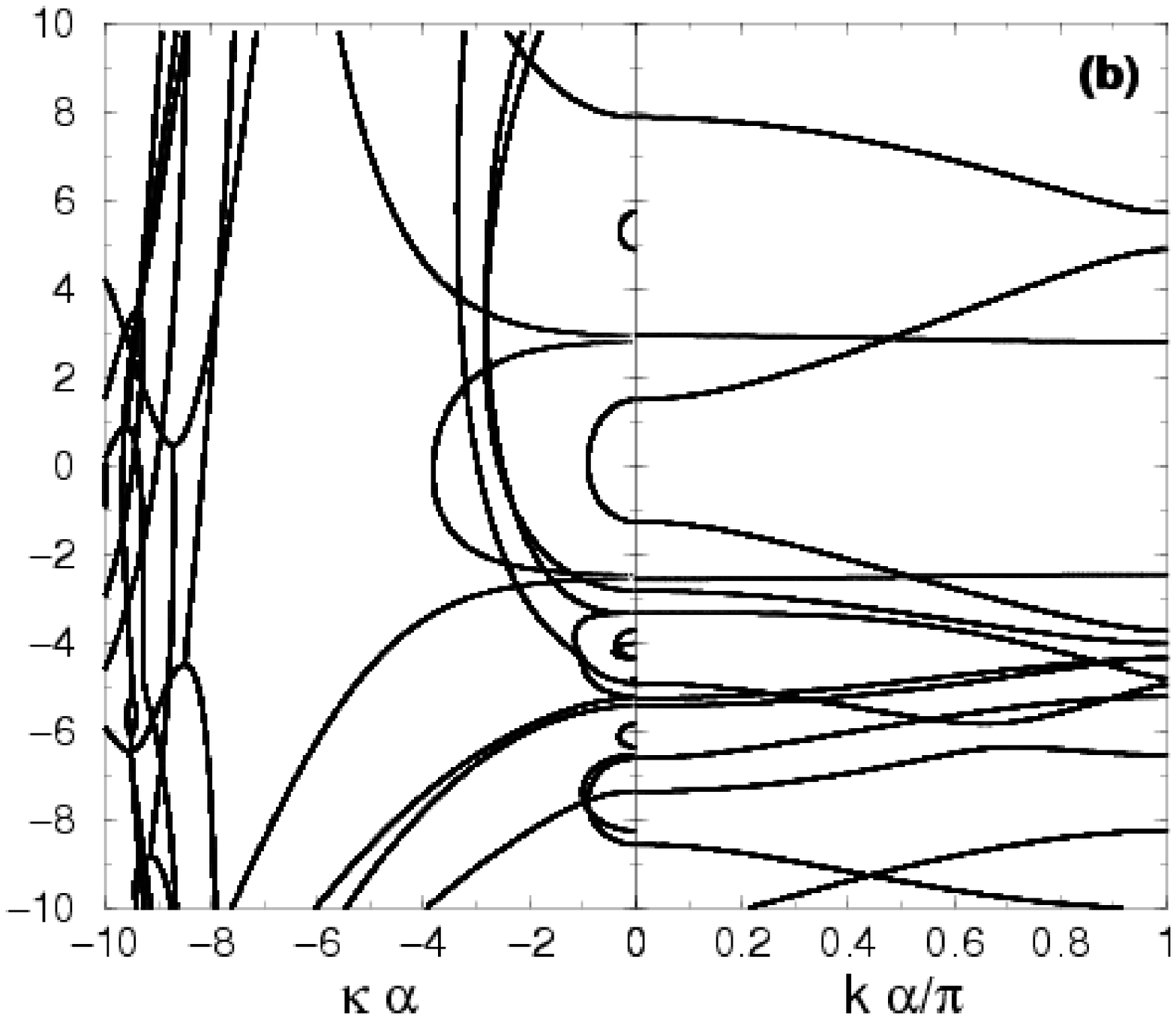,width=14pc,height=14pc}}
\end{minipage}

\begin{minipage}{0.5pc}
\begin{rotate}{90}
E-E${\rm_F}$
\end{rotate}
\end{minipage}
\begin{minipage}{15pc}
\centerline{\epsfig{figure=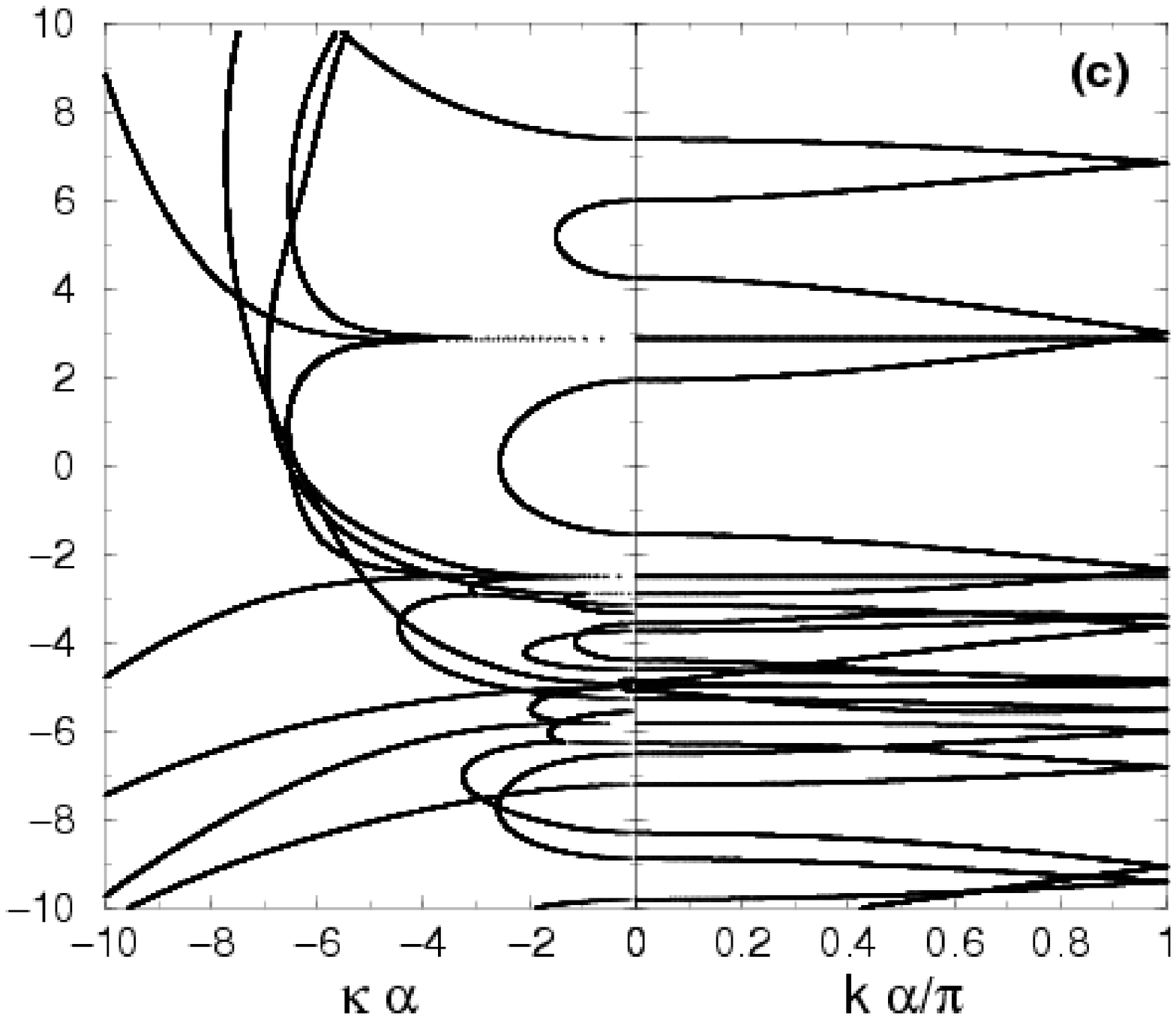,width=14pc,height=14pc}}
\end{minipage}

\caption{Electronic band structure for (a) PPE, (b) planar PPP, and (c)
non-planar PPP (tetrahedral angle at $45^{\circ}$).
The right panel indicates the band structure
for real $k$ in the first Brillouin zone, and the left panel
the complex-band structure.}
\label{f4}
\end{figure}

In all cases, one has to solve the Schr\"{o}dinger equation in a periodic
potential, which in our tight-binding approximation reads
\begin{equation}\begin{split}
&\left(\mathbf H_{m,m}^{TB}+\mathbf H_{m,m+1}^{TB}e^{ika}+\mathbf H_{m-1,m}^{TB}e^{-ika} \right)
\phi_k= \\ &E_k
\left(\mathbf S_{m,m}^{TB}+\mathbf S_{m,m+1}^{TB}e^{ika}+\mathbf S_{m-1,m}^{TB}e^{-ika} \right)\phi_k.
\label{eq9}
\end{split}\end{equation}
Here, the matrices $\mathbf H_{i,j}^{TB}$ and $\mathbf S_{i,j}^{TB}$ are defined
as containing the Hamiltonian and overlap
matrix elements, respectively, between the $i$th and $j$th unit cells.
$a$ is the lattice constant.

To obtain the complex-band structure, one
follows the opposite procedure of typical band structure calculations
in which the solution of the eigenvalue problem of the
Hamiltonian yields the energy spectrum for any real $k$.
The spectrum of $k$ vectors associated with any real energy $E$ is
sought. Therefore, it is useful to
transform Eq. \ref{eq9} from an eigenvalue problem for $E_k$ to an
eigenvalue problem for $\lambda\equiv e^{ika}$. This problem resembles
finding the eigenvalues of the transfer matrix. It may be
cast in a generalised eigenvalue problem via
\begin{equation}\begin{split}
&\left(\begin{array}{cc}
\mathbf H_{m,m}-E\mathbf S_{m,m} & \mathbf H_{m-1,m}-E \mathbf S_{m-1,m} \\
\mathbf I & \mathbf 0 \end{array} \right)
\tilde{\phi}_\lambda= \\
&\quad\lambda\left(\begin{array}{cc}
-(\mathbf H_{m,m+1}-E \mathbf S_{m,m+1}) & \mathbf 0 \\
\mathbf 0 & \mathbf I \end{array} \right)\tilde{\phi}_\lambda,
\label{eq10}
\end{split}\end{equation}
which in contrast to usual formulations,
can be used even for singular $\mathbf H_{i,j}-E \mathbf S_{i,j}$,
as it often happens when a localised basis set is employed.
When $\left |\lambda \right|=1$, $k$ is real; otherwise $k$ is complex and
we have exponentially decaying or growing solutions, which come in
conjugate pairs.

The complete band structures of perfect PPE, planar PPP, and non-planar PPP
are calculated using Eq. \ref{eq9} and Eq. \ref{eq10}, and they
are shown in Fig. \ref{f4}(a), \ref{f4}(b), and \ref{f4}(c), respectively.
At the right panel we present the conventional band structure.
The spectrum of the imaginary part of the
complex $k$ solutions (negative branch) is plotted at the left panel .
Our calculated band structures compare well with those reported
for either minimal \cite{JCP04TS,PRB02TS} or plane wave
\cite{PRB02TS,PRB00KMS} basis sets within other
DFT methods in the LDA, both in the real axis and in the complex plane.
As expected from studies of the analytic properties of
the energy spectrum in the complex plane,
the complex-band structure exhibits the connection of real bands
between local extrema. Of particular importance for the off-resonant
electron transport phenomena we discuss in the next section, is
the, so-called, real line which joins the edges of the
valence and conduction bands which develop around $E_F$.
It evidently corresponds to the smallest decay of wavefunctions into
the molecular region since $\left |\kappa \right|$ is minimum, and
it is expected to dominate in tunnelling processes within $E_g$.

\section{Conductance properties}
\label{sec:S4}

\subsection{General Features}
\label{sec:S4A}
\begin{figure}[h]
\centerline{\epsfig{figure=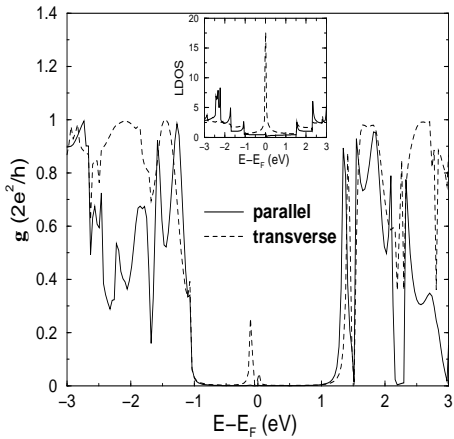,width=14pc,height=14pc}}
\caption{Conductance spectrum for an OPE wire with
length $N=2$. The solid line corresponds to parallel connection, and
the dashed line corresponds to transverse connection (see Fig. \ref{f1}).
The inset shows the local DOS at the two carbon apex-sites
for parallel (solid line) and transverse configuration (dashed line).}
\label{f5}
\end{figure}

The local microstructure of the molecule/electrode interfacial
contact, especially for low-dimensional electrodes \cite{PRB01FCR,CP02CFR},
has been
shown to crucially determine the conductance properties of a
molecular junction. This is also seen in the transport properties
of structures such as those of Fig. \ref{f1}. To demonstrate the
bare effect, we take the parallel and transverse geometries
of a diphenyl-ethynyl bridge. As shown in Fig. \ref{f5},
a clear distinction of the conductance spectrum is evident.
Compared to previous studies \cite{PRB01KB,EPL03GFR,PRL03EK},
a remarkable fact is that despite the carbon-carbon covalent bond
at the interfaces and the close conformational similarity,
there is a prominent dissimilarity in the conductance.
The latter can be traced to distinct features in the local
DOS for the zigzag and the armchair edge topologies (inset of Fig. \ref{f5}).
An analogue of this effect arising from the molecular bridge
contributions to the electronic states at $E_F$ was observed in Refs.
\cite{JPCB03TY,CPC03GGS}.

In particular, the presence of the zigzag edge state has a profound influence.
The conductance through the oligomer is non-negligible within
the HOMO-LUMO gap, exhibiting a double resonant peak around $E_F$.
For the parallel structure, on the other hand, the conductance is almost
vanishing inside $E_g^M$. The observed conductance resonances are in accordance with
previous results showing that due to a large
local DOS on the metal side for energies inside the HOMO-LUMO gap,
an additional resonant channel becomes available for
through-bond conduction \cite{EPL03GFR,MS04FGR}. Indeed, the
inset of Fig. \ref{f5} presents the local DOS at edge carbon-atoms
of an isolated semi-infinite zigzag ribbon. The dashed line is
the local DOS at a carbon atom (saturated by hydrogen) at the zigzag edge,
whereas the solid line is for
an atom located at an armchair mid-point of the terminated ribbon.
The localised state of zigzag edge sites is featured in their local DOS
by the characteristic spike, but is missing for armchair
edge sites.
\begin{figure}[h]
\centerline{\epsfig{figure=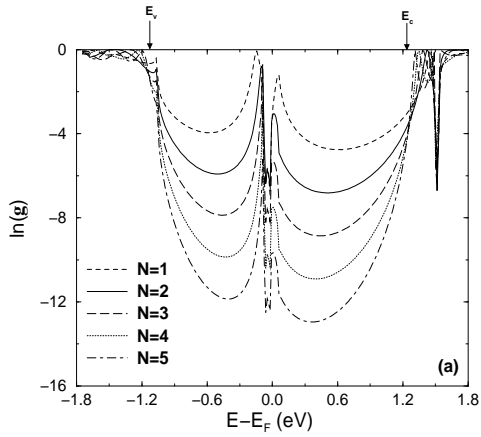,width=14pc,height=14pc}}
\centerline{\epsfig{figure=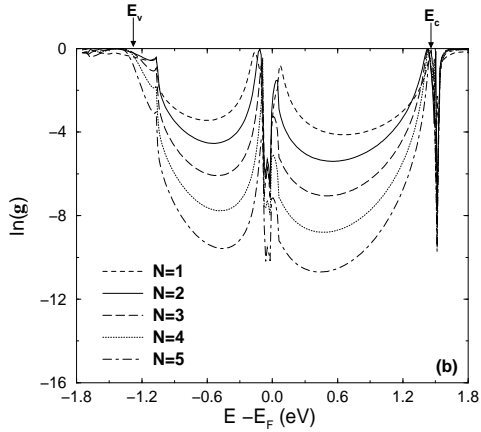,width=14pc,height=14pc}}
\centerline{\epsfig{figure=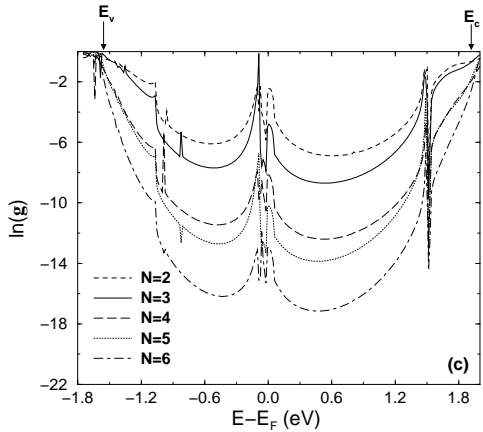,width=14pc,height=14pc}}
\caption{Logarithm of ${\mathrm g}$ versus energy $E$, for wires of
various lengths $N$. (a) OPE, with  valence and conduction band energies
$E_v=-1.12$eV and $E_c=1.23$eV, respectively. (b) planar OPP, with
$E_v=-1.27$eV and $E_c=1.48$eV. (c) non-planar OPP, with $E_v=-1.55$eV and
$E_c=1.91$eV.}
\label{f6}
\end{figure}

To discuss generic properties of off-resonant electron
transport properties within $E_g$, we focus on
junctions of the transverse connection, which {\it exhibit} additional
conductance resonances within $E_g^M$ due to the topology of
the electrode/molecule interface ('worst case scenario').
In Fig. \ref{f6} we present the logarithm of the conductance spectrum
of such carbon junctions, for various lengths $N$ of the oligomers.
Plot (a) corresponds to OPE, with valence and conduction band energies
of the corresponding polymer wire $E_v=-1.12$eV and $E_c=1.23$eV,
respectively (see Sec. \ref{sec:S3B} and Fig. \ref{f4}) .
Plot (b) is for planar OPP, with $E_v=-1.27$eV and $E_c=1.48$eV, and
plot (c) is for non-planar OPP, with $E_v=-1.55$eV and $E_c=1.91$eV. As
it has been expected from the above discussion, the conductance
exhibits resonances around $E_F$, which are relatively high for small $N$.
Even though ${\mathrm g(E)}$ decreases as $N$ becomes larger,
these peaks pertain for all wire lengths. Different oligomers,
thus different electronic configurations of the
molecule and different coupling to the electrodes, do not obscure the
appearance of these conductance resonances, indicating once again
that they arise solely from the topology of the electrode contacting
surface.
\begin{figure}[h]
\centerline{\epsfig{figure=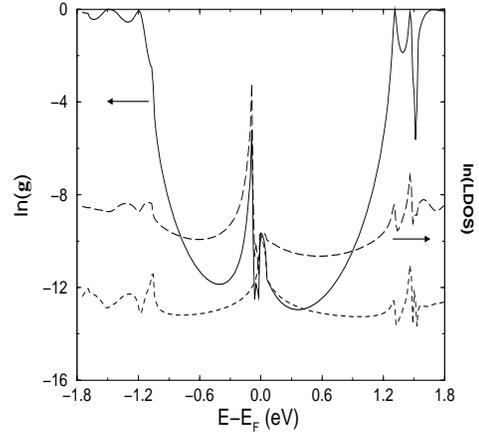,width=14pc,height=14pc}}
\caption{The left $y$-axis corresponds to $\ln({\mathrm g})$, and the
right $y$-axis to the logarithm of the local DOS at the carbon atomic
contacts, versus $E$, for OPE with $N=5$. The dotted line is the DOS at the
contact point between the lower electrode and OPE, and the dashed line
is for the DOS at the upper one (see Fig. \ref{f1}).}
\label{f7}
\end{figure}

The existence of both peaks in Fig. \ref{f6} can be
understood by inspection of Fig. \ref{f7}. There,
$\ln({\mathrm g})$ of an OPE wire of length $N=5$ is plotted together
with the logarithm of the local DOS at the lower (dotted line) and
upper (dashed line) electrode apex atoms in contact with the molecule,
as a function of the energy. Each of the local DOS at the two contacts
has a double peak inside the gap. The energies at which the two conductance
peaks appear correspond to the points on the energy range within the
gap for which there is large local DOS. This result can be roughly
understood from the $\pi$-orbital approximation, in which the transmission
is related to the local DOS $\nu_{L/U}$ at the lower/upper atomic contacts
through the relation
$T(E)\propto\nu_L(E)\nu_U(E)\left|G_{1N}(E)\right|^2$, $G_{1N}$
being the propagator between the first and the last sites of the molecule
\cite{CP02CFR,JCP94MKR,SSC98OKM}.
\begin{figure}[h]
\centerline{\epsfig{figure=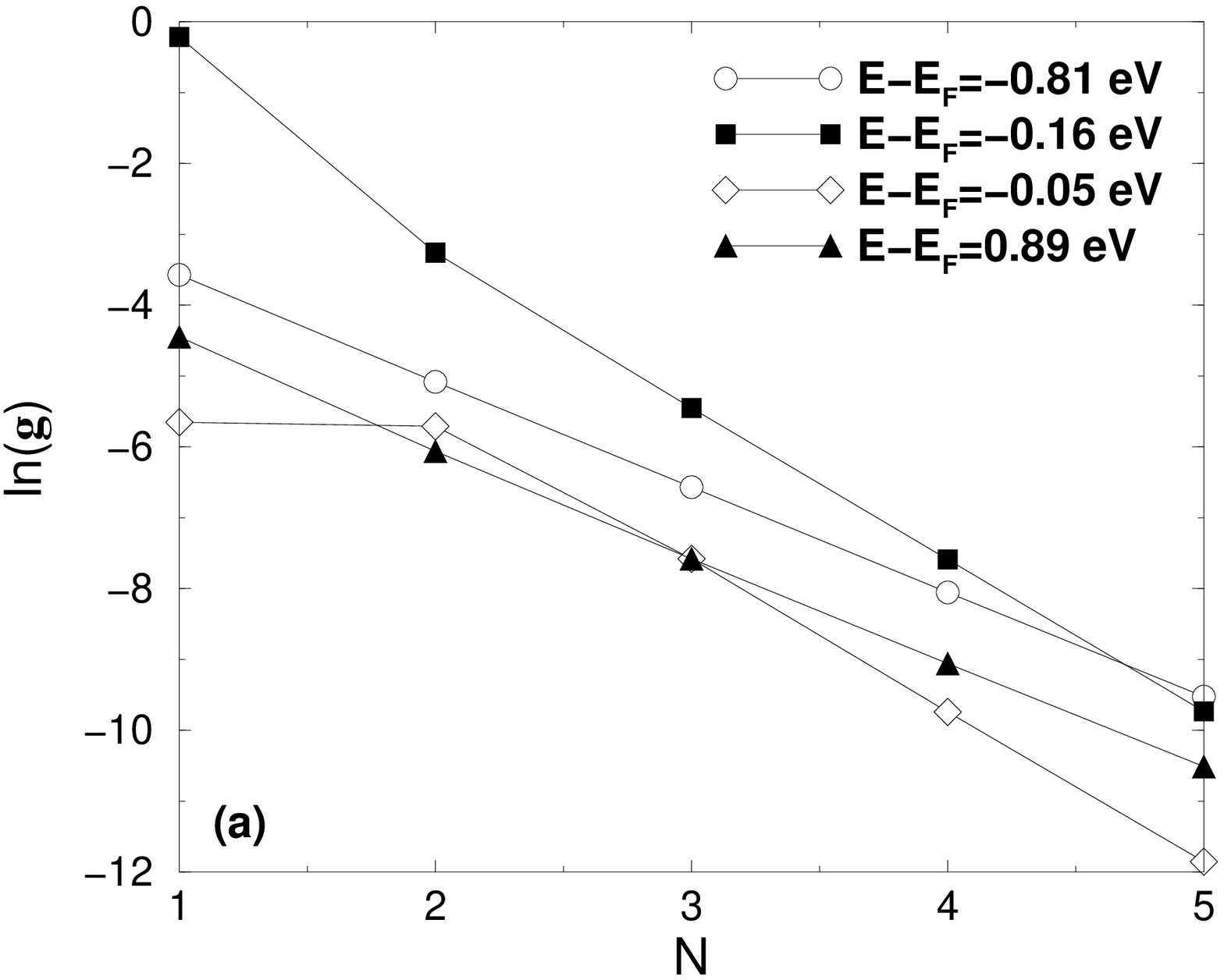,width=14pc,height=14pc}}
\centerline{\epsfig{figure=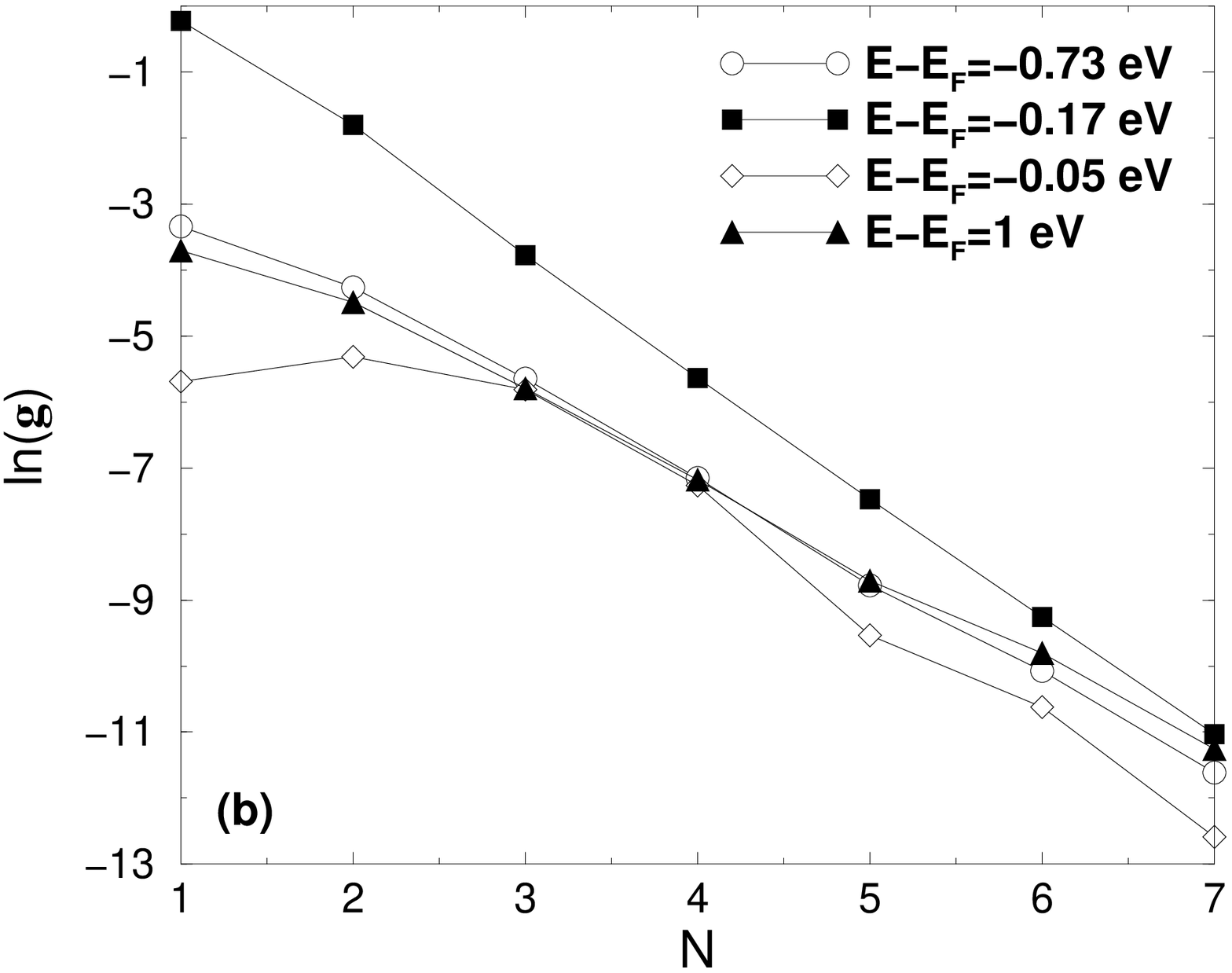,width=14pc,height=14pc}}
\centerline{\epsfig{figure=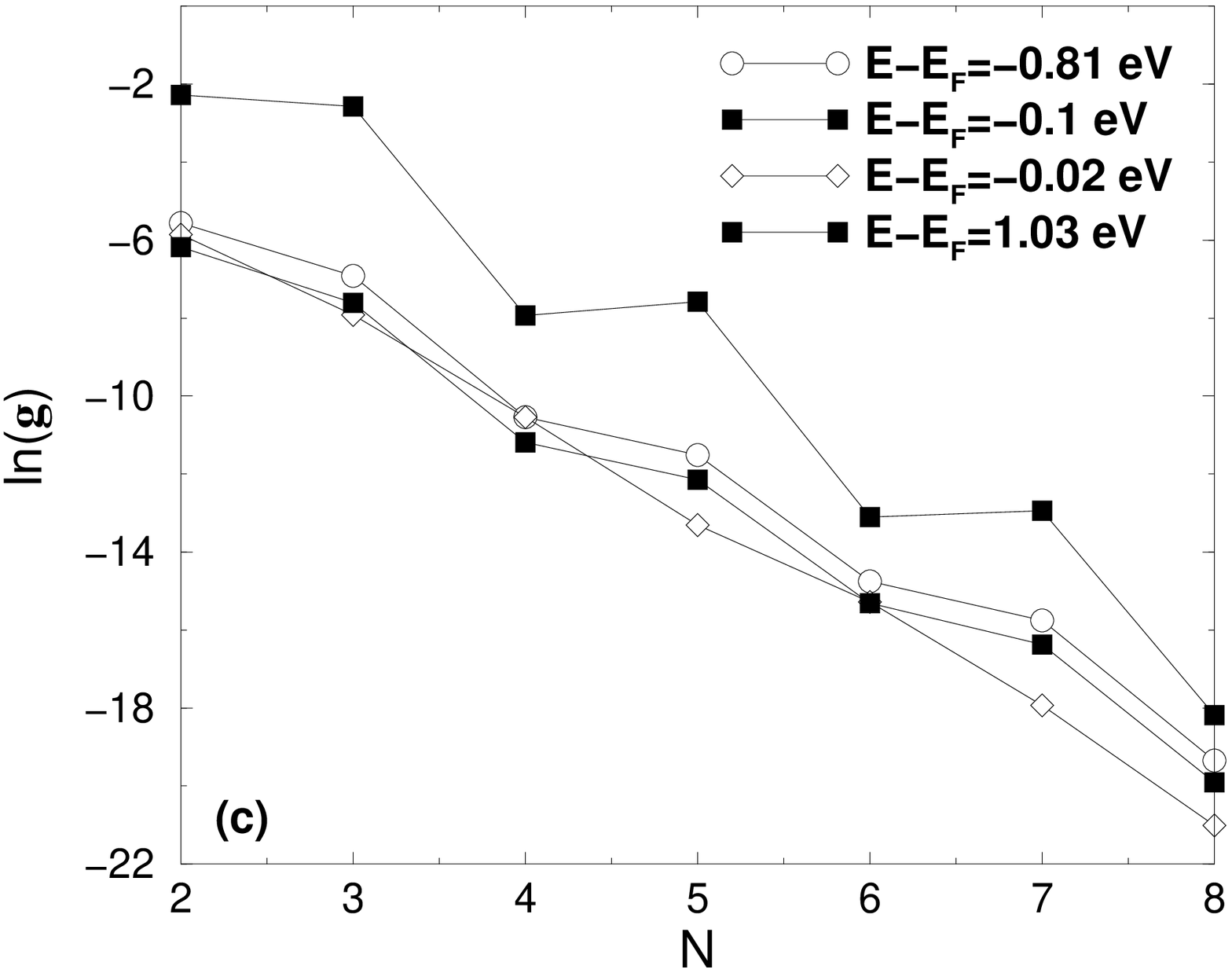,width=14pc,height=14pc}}
\caption{$\ln({\mathrm g})$ versus $N$ for various energies within the
HOMO-LUMO gap. (a) OPE, (b) planar OPP, and (c) non-planar OPP.}
\label{f8}
\end{figure}

From Fig. \ref{f6} we see that the conductance decreases with the
length of the molecule. Examining this length dependence in more detail,
$\ln({\mathrm g})$ is plotted as a function of $N$ for selected energies inside
the gap, in Fig. \ref{f8}.
Graph (a) is for OPE, graph (b) is for planar OPP, and graph (c) is for
non-planar OPP. The energy values of the curves defined by circle and triangle
values reside in the left and right side, respectively,
of the edge states spectrum. Plots indicated by the squares
correspond to one of the two conductance resonances, and the remaining curves
(circles) are for an energy between the two peaks.
In all three graphs, we evidence three types of curve.

In Figs. \ref{f8}(a) and \ref{f8}(b) and for energies away from those of
the edge states, $\ln({\mathrm g})$ falls linearly with $N$.
This is reminiscent of the tunnelling behaviour, Eq. \ref{eq1}. In contrast,
for energies at the vicinity of the edge state resonances,
an anomalous length dependence appears for very small $N \leq 3$, but as
$N$ increases, a linear decrease is also attained. We attribute the former
to slight positional shifts of the resonant energies which are
evident in the plots as $N$ varies from  1 to approximately 3.
After a complete analysis of the conductance spectrum within $E_g$,
we conclude that the exponential law of Eq. \ref{eq1} holds generally
for the entire energy range within  $E_g$ and for ultrashort molecular
junctions. The observed few deviations are specific to the appearance of
"unconventional" resonant channels induced by the electrodes.

For the non-planar OPP (Fig. \ref{f8}(c)), it appears that
$\ln({\mathrm g})$ decreases with $N$ in an oscillating rather
than a linear fashion. A closer look, however, reveals that one should
distinguish between the conductance for even and odd $N$.
We should bear in mind the difference in the interfacial contact geometry
of the last benzene ring of the oligomer. For odd $N$,
the benzene ring is at the same plane with that of the
ribbon, whereas for even $N$ it is positioned at an angle $\theta$
away from the plane. This distinction results in different coupling
with the electrode, and therefore, in a different contact conductance
as we analyse in Sec. \ref{sec:S4B}. A clear difference is also reflected
in the structure of the conductance spectrum around $E_F$ in Fig. \ref{f8}(c).
It follows, however, that the exponential law applies for the
two cases separately, with the same inverse decay length. Another viewpoint
would be to consider an increase of the molecular length to be given
in units of the unit cell of the non-planar PPP.

\subsection{Inverse Decay Length}
\label{sec:S4B}
\begin{figure}[h]
\centerline{\epsfig{figure=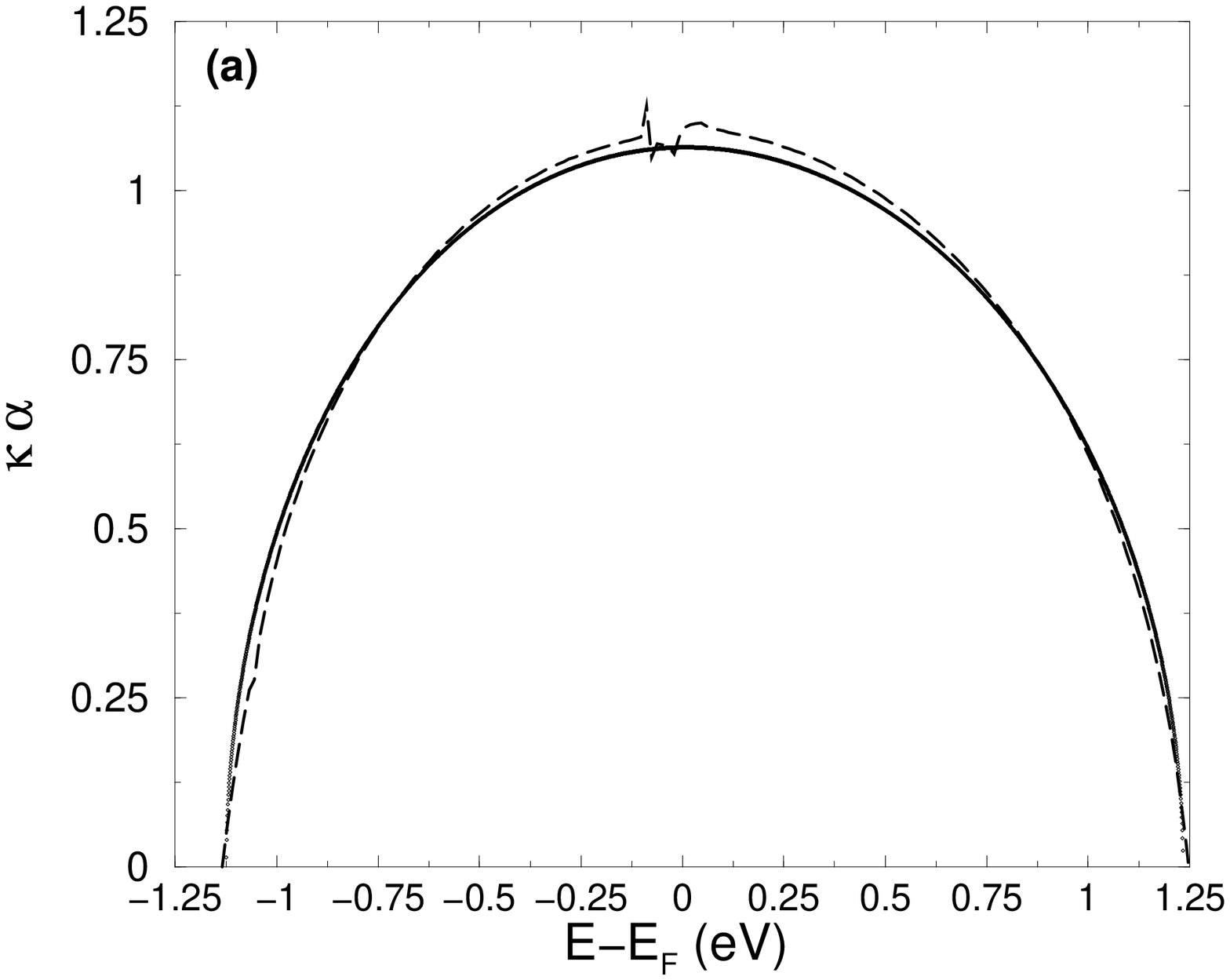,width=14pc,height=14pc}}
\centerline{\epsfig{figure=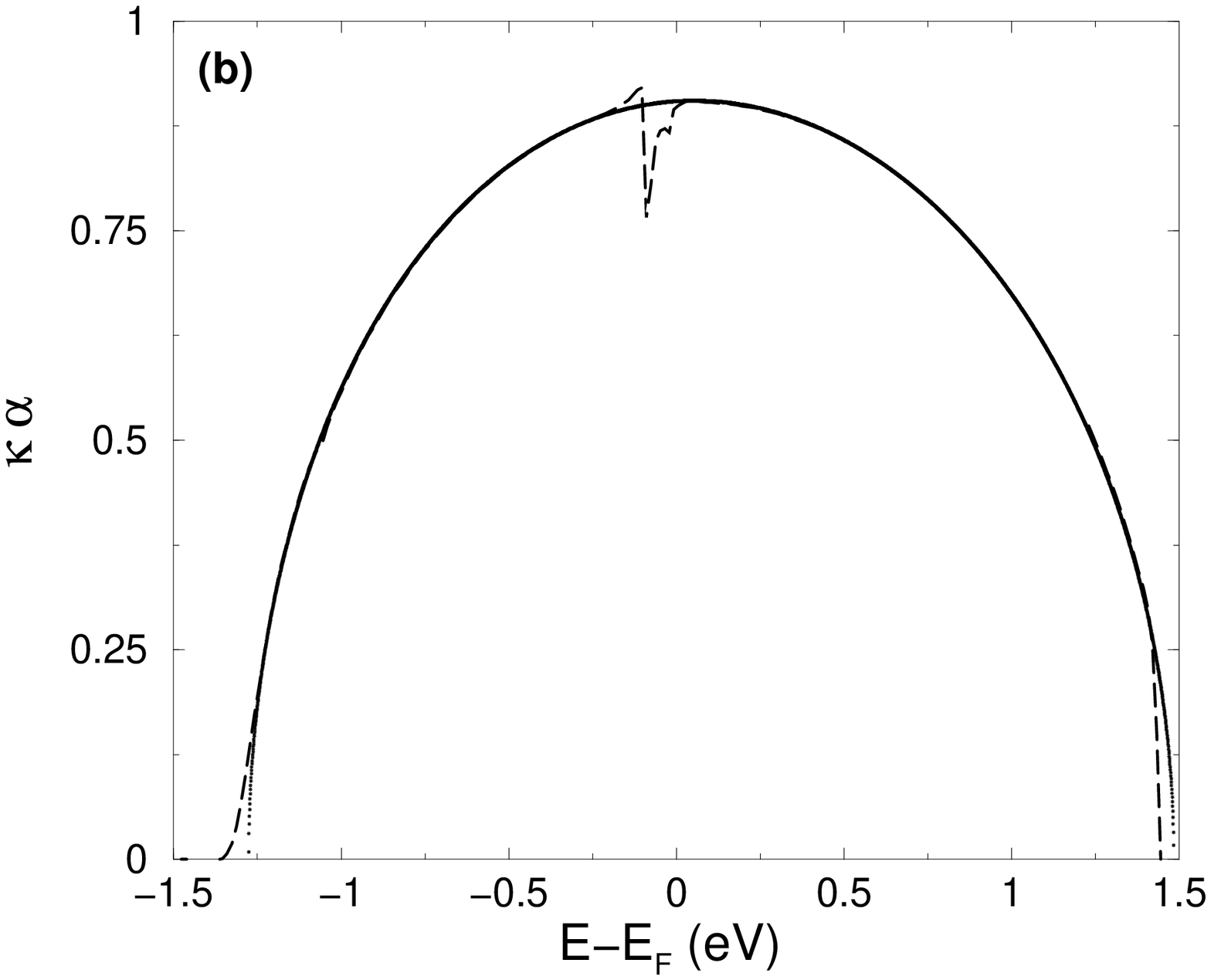,width=14pc,height=14pc}}
\caption{Complex-band structure ($E_c-E_v$ real line) plotted with
a solid curve and tunnelling inverse decay lengths for (a) PPP and OPE,
(b) planar PPP and OPP, respectively.}
\label{f9}
\end{figure}
In the previous Section we established Eq. \ref{eq1}
for virtually all molecular lengths and for any energy in the regime
of interest. Here, we identify the damping factor of the tunnelling 
conductance of an oligomer junction with the complex-band structure of
the corresponding idealised polymer wire discussed in Sec. \ref{sec:S3B}.
Consistent with the assumption that, to a first approximation,
evanescent states in the junction would coincide with those of the
idealised long wire and that those with the smallest
$\left|\kappa\right|$ are dominating the tunnelling process,
in Fig. \ref{f9} we isolate the
positive branch of the real line $E_c-E_v$ (solid line). The latter
equals minus the negative branch. Together we plot
the inverse decay length $\beta$
divided by a factor two (dashed line), which accounts
for transport coefficients relating to probability amplitudes
rather than wavefunctions. $\beta$ is numerically obtained
from a least-squares fit for the determination of the slope of the
lines of Fig. \ref{f8}. Graph (a) corresponds to OPE and
graph (b) to planar OPP. For energies at which the conductance shows
anomalous length dependence, we use only data for $N>2$.
In both graphs (a) and (b), there is a remarkable
agreement
%between that $\kappa$ with the largest imaginary part within $E_g$ and $\beta/2$,
for the entire energy window.
There are no deviations from the relation $\beta=2\left|\kappa(E)\right|$
apart from energies at which the graphitic ribbons exhibit a large local DOS.
\begin{figure}[h]
\centerline{\epsfig{figure=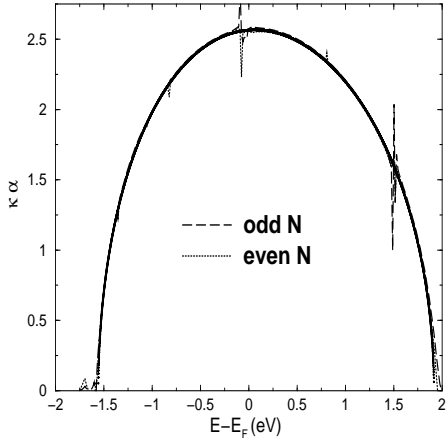,width=14pc,height=14pc}}
\caption{Complex-band structure ($E_c-E_v$ real line) plotted with a solid
curve and tunnelling inverse decay lengths for non-planar PPP and OPP,
respectively. The dashed line corresponds to $\beta/2$ from OPP with
odd $N$ and the dotted one is for OPP with even $N$.}
\label{f10}
\end{figure}

In the discussion of Fig. \ref{f8}, we mentioned that for
non-planar OPP wires the parity of the length is an
important element for the exact conductance spectrum, but
that the inverse decay lengths apparently coincide for even and odd $N$.
This is evident in Fig. \ref{f8} where the damping factor divided by two
is plotted for both cases and is compared to the complex-band structure with
the smallest $\left|\kappa(E)\right|$. The damping factors
for wires with odd (dashed line) and even $N$ (dotted line)
are almost identical to each other. Furthermore, they fall on top of
the positive branch of $\kappa$ which derives from the conduction and
valence bands of non-planar PPP. Once again there is absolute
quantitative agreement
%of the conductance damping factor to this real line of the complex-band structure
apart from energy points around
singularities in the local DOS (see Fig. \ref{f3}).
It is worth mentioning that all results for the extracted tunnelling
decay lengths in oligomer junctions are obtained for the optimised
geometries, whereas, the complex-band structure is
that of an ideal polymer wire. This shows the robustness of the mapping
to small perturbations on atomic positions.
\begin{figure}[h]
\centerline{\epsfig{figure=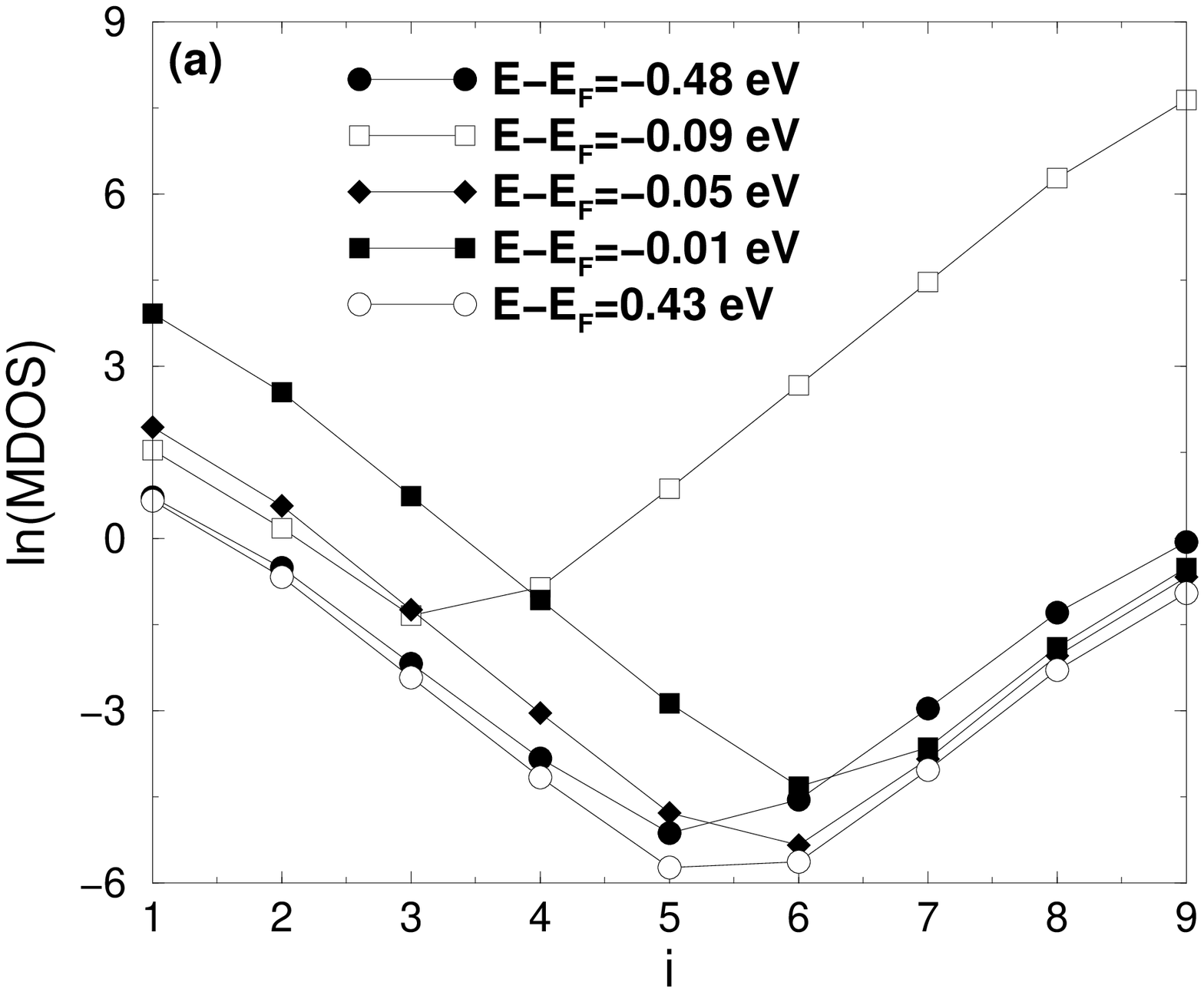,width=14pc,height=14pc}}
\centerline{\epsfig{figure=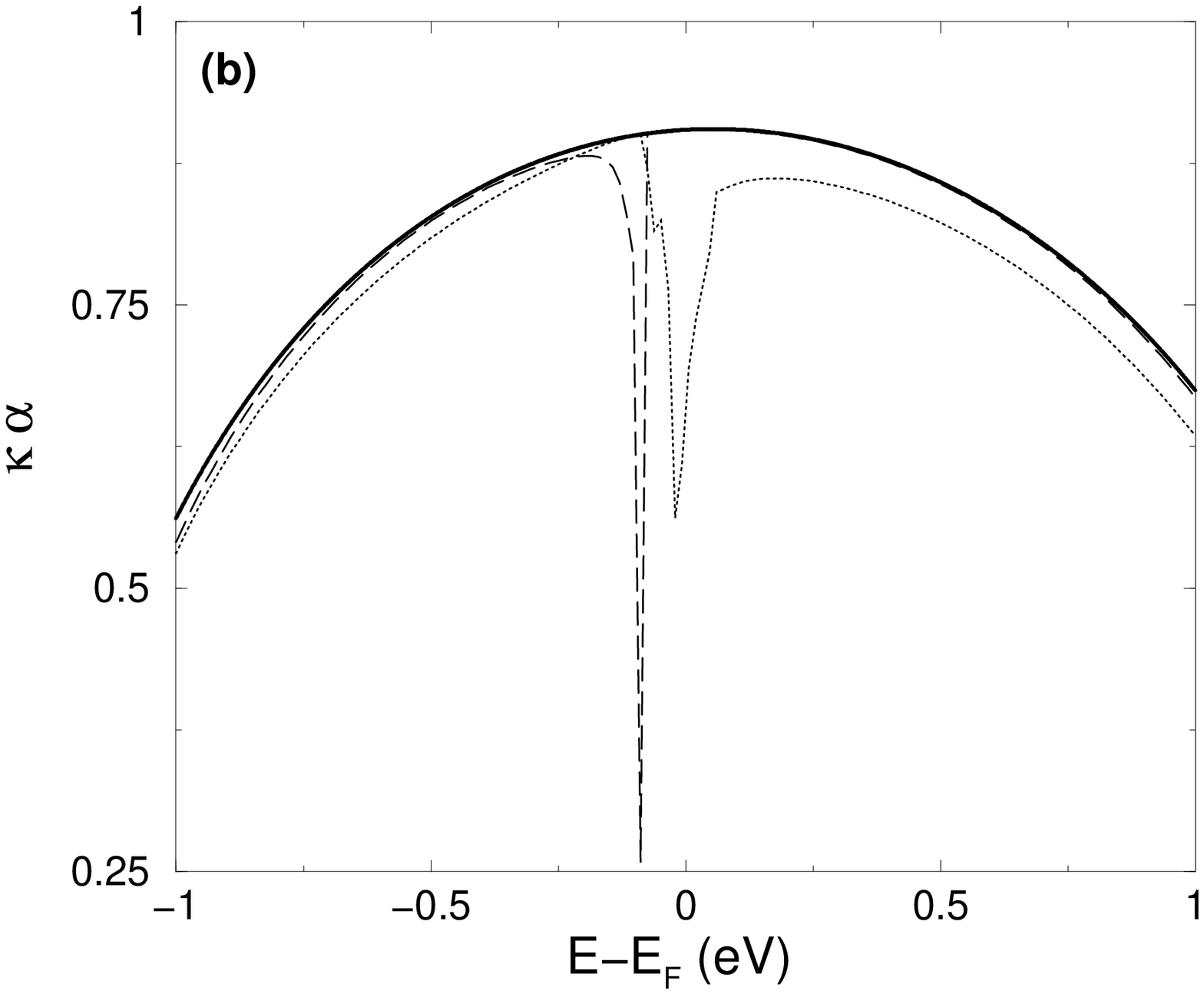,width=14pc,height=14pc}}
\caption{(a) Logarithm of DOS on a planar OPP wire of length $N=9$ as a
function of the monomer unit number $i$, for selected energies in
the HOMO-LUMO gap. (b) Damping factor $\beta_{MDOS}$ for the wire of (a), from
least-squares fitting of $i=2, 3, 4$ (dashed line), and of $i=6, 7, 8$
(dotted line).}
\label{f11}
\end{figure}

According to the arguments of Sec.\ref{sec:S3B} on the wavefunction
matching between the metal and the molecular wire, 
the density of states inside the molecule should also
reflect properties of decaying states. Namely,
its projection to an exponential should be possible,
at least away from the electrode/wire interface.
We examined this approach by calculating the projected density of states
within a unit cell (MDOS) of a planar OPP wire of fixed length.
In Fig. \ref{f11}(a), we plot its logarithm as a function of the
monomer units $i$ which make up $N=9$,
for energies chosen inside the gap. We see
that the logarithm of MDOS falls off almost linearly with $i$ until roughly
half of the molecular length is reached, and then $\ln$(MDOS) increases,
again in a linear fashion. Loosely speaking, the decreasing part of
each curve corresponds to the decaying metal-induced gap state
from the lower electrode whereas the increasing part comes from
the gap state of the upper electrode. This holds for relatively
long wires as the one presented. For shorter molecular bridges,
none of the contributions from each side is dominant. No
apparent linearity as in Fig. \ref{f11}(a) is observed.

%The asymmetry between the left and right parts of MDOS is due to differences on the local interfacial geometry after optimisation.

To make this analysis quantitative, we calculate a damping factor
$\beta_{\rm MDOS}$ for the density of states,
presented in Fig. \ref{f11}(b). The solid
line is again the smallest $\left| \kappa \right|$
from the complex band structure of planar PPP (Fig. \ref{f9}(b)).
The dashed and dotted lines represent $\beta_{\rm MDOS}/2$
from a least-squares fit to $\ln$(MDOS) for $i=2$ to $4$ and $i=6$ to $8$,
respectively. Both these curves agree reasonably well with
$\left| \kappa \right|$, corroborating the picture of metal-induced gap
states. We interpret the asymmetry between the left and right parts of MDOS in
Fig. \ref{f11}(a) and (b) as differences on the atomic interfacial contact
after optimisation. For example, the erroneous divergence
of $\beta_{\rm MDOS}/2$ around $E_F$ emerges at different energies
in Fig. \ref{f11}(b) for gap states arising from the lower and
the upper electrode, with the former seeing the large DOS from the
upper electrode, and the latter seeing the large DOS from the lower
electrode (see Fig. \ref{f7} for comparison).

We briefly comment on the correspondence between
the inverse decay length obtained from the molecular density of states
and the transport calculations. It is evident that the exponential
decay of wavefunctions across the molecular bridge as quantified
by the complex-band structure governs both $\beta_{\rm MDOS}$ and $\beta$.
However, the spectral properties are determined by the diagonal Green
function $G_{ii}$. Transport is ultimately related to the propagator $G_{ij}$
between unit cells, i.e,  $i\neq j$. As suggested by the above results
and the analytic properties of simple $\pi$-orbital models
the convergence of the latter to the long oligomer limit is much
faster, enabling an analysis of the tunnelling regime via the complex $k$.
\begin{figure}[h]
\centerline{\epsfig{figure=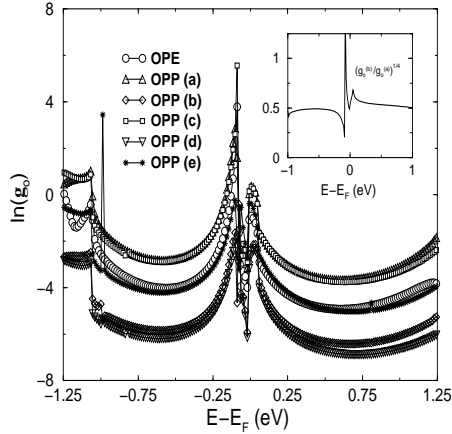,width=14pc,height=14pc}}
\caption{Logarithm of contact conductance ${\mathrm g}_o$ versus energy
for all oligomer junctions under study.}
\label{f12}
\end{figure}

\subsection{Contact Conductance}
\label{sec:S4C}

The results of the previous Section elucidate that the
inverse decay length $\beta$ of the conductance in the tunnelling
regime is an intrinsic property of the oligomer embedded in the
nanojunction. By contrast the prefactor ${\mathrm g}_o$ of
Eq. \ref{eq1} is largely determined by the realised
interfacial contact. To investigate the so-called contact conductance,
in Fig. \ref{f12} we present its logarithm versus energy. The curve
denoted as OPE corresponds to the contact conductance of
similar (or longer) molecular junctions to that of Fig. \ref{f1}.
Those curves indicated by OPP refer to the
molecular wires drawn in Fig. \ref{f2}. All data are obtained
after ensuring that the exponential decay of Eq. \ref{eq1}
is fulfilled as explained in Sec. \ref{sec:S3B}.
Apparently, ${\mathrm g}_o$ depends on the exact atomic contact
(whether phenyl or ethynyl group to graphene) as well as the orientation
of the phenyl rings with respect to the graphitic ribbons.

It is possible to identify pairs of similar
contact conductances from Fig. \ref{f12}:
i) OPE with OPP (e), ii) OPP (a) with OPP (c), and iii) OPP (b) with
OPP (d). Although the origin of the first pair is probably fortuitous,
pairs ii) and iii) are expected from simple arguments. Moreover,
the smaller coupling of the ethynyl-group with respect to
the phenyl rings should give rise to the observed reduced ${\mathrm g}_o$
of OPE compared to that of planar OPP. The different magnitude of
${\mathrm g}_o$ between non-planar oligo-para-phenylene
for odd, OPP (c), and even, OPP (e), number of unit cells in the
wire, expresses the importance of the length parity in
experiments. This is in agreement with our previous discussion.
The increased resistance for non-planar alignments of the
phenyl end-rings with respect to the electrodes, e.g.,
OPP (b), OPP (d) and OPP (e), may be explained as a break-up of
the conjugation.

In addition, we estimate the contact conductance ratio between the
parallel aligned and twisted phenyl end-ring geometries.
We assume, as previously, only $\pi$-electrons in an orthogonal basis set.
This simple model yields ${\mathrm g}_o (E)= \Gamma^2_L\Gamma^2_U f(E)$,
where, $\Gamma_{L/U}$ corresponds to Hamiltonian matrix elements
between the apex carbon atoms located at the wire and the graphene edges.
Furthermore, we may write
$({\mathrm g}_o^{(b)}/{\mathrm g}_o^{(a)})^{1/4} \approx  \Gamma^{45^\circ}/\Gamma^{0^\circ}$,
which yields $cos(45^\circ) \approx 0.7$. Considering the oversimplification,
this agrees well with the accumulated data (inset of Fig. \ref{f12}),
which give a value around 0.5, apart from the region where the local
DOS on the metal side is singular.

Finally, we note the non-trivial energy dependence of the contact conductance,
which may hinder simple estimations based on the knowledge of $\beta$ as
those suggested in Ref. \cite{JCP04TS}. In all our studied cases
the contact conductance follows roughly the behaviour of the local
DOS at the electrode atomic contacts, with a pronounced peak around
the Fermi energy, underlining the importance of the electronic structure
of the electrodes. This follows from $f(E)$ being formally decomposed
to $\nu_L(E)\nu_U(E)X_M(E)$ in the aforementioned simple model, i.e.,
being proportional to the product of the metal local DOS, $\nu_L(E)\nu_U(E)$
and a molecular factor $X_M(E)$, which allows for resonant tunnelling
at molecular orbital energies. Although $f(E)$ may be a slowly varying function
of energy for leads with a seamless local DOS, especially deep in the
tunnelling regime, it is not a priori clear that the off-resonant contact
conductance is energy independent and can be determined from the
on-resonance value \cite{JCP04TS}. The latter would imply equal contact
conductances for perfectly symmetric structures based on OPP (a) and OPP
(b) of Fig. \ref{f2} although these are at least an order of magnitude
different ($\sim 0.5^{-4}$). Clearly, more detailed investigations in
diverse molecular junctions are necessary.

\section{Concluding Remarks}
\label{sec:S6}

We studied the generic properties of off-resonant electron transport
across molecular junctions of current experimental interest,
formed from oligomers bridging graphitic electrodes.
To this end, we used a comprehensive analysis of the electronic
structure of the various components in a variety of possible
configurations. We showed how the conductance magnitude is crucially
determined by the electrode spectral properties, and the exact topological and
chemical realisation of the interfacial contact. Most importantly, as
a function of the molecular length the tunnelling conductance
is almost always an exponentially decreasing function with the
inverse decay length given intrinsically by the complex-band structure
of the corresponding idealised polymer chain. The properties and accuracy
of the latter can be studied independently (cf \cite{JCP03PSS}).
Once the Fermi energy is also located with confidence,
the damping factor is theoretically extracted.
This fact has great implications on comparing theoretical and experimental
results. A current issue is the discrepancy between those. However,
the exact contact geometry, which may have dramatic influence, is largely
unknown. Therefore, direct comparison of decay constants in well defined
geometries, e.g., in self-assembled monolayers, may shed more light in
our understanding of electron transport across molecular junctions.

\section*{Acknowledgements}
The authors appreciate the kind permission of Marcus Elstner
of the TB-DFT team to use his version of the code.
AK would like to thank the {\it Alexander von Humboldt Stiftung}.
Great part of this work has been completed while GF was at Universit\"{a}t
Regensburg and supported by the {\it Graduiertenkolleg Nichtlinearit\"{a}t und
Nichtgleichgewicht in Kondensierter Materie}. GF acknowledges current funding
by the ATOM CAD project within the {\it Sixth Framework Programme} of EU.


\begin{thebibliography}{99}
\section*{References}

%REVIEWS
%1
\bibitem{Nat00JGA}
C. Joachim, J.K. Gimzewski, and A. Aviram, Nature {\bf 408}, 541 (2000)

%2
\bibitem{Sci03NR}
A. Nitzan and M.A. Ratner, Science {\bf 300}, 1383 (2003)

%3
\bibitem{JCM03S}
T. Seideman, J. Phys.: Condens. Matter {\bf 15}, R521 (2003)

%HISTORIC
%4
\bibitem{CPL74AR}
A. Aviram and M.A. Ratner, Chem. Phys. Lett. {\bf 29}, 277 (1974)

%EXPERIMENT
%5
\bibitem{Sci97RZM} M.A. Reed, C. Zhou, C.J. Muller {\it et al}
Science {\bf 278}, 252 (1997)

%6
\bibitem{JCP98TDH}
W. Tian,  S. Datta, S. Hong {\it et al}, J. Chem. Phys. {\bf 109},
2874 (1998)

%7
\bibitem{Sci01CPZ}
X.D. Cui, A. Primak, X. Zarate {\it et al}, Science {\bf 294}, 571 (2001)

%8
\bibitem{Nat02SNU}
R.H.M. Smit, Y. Noat, C. Untiedt {\it et al},
Nature {\bf 419}, 906 (2002)

%9
\bibitem{APL03RBW}
J. Reichert, D. Beckmann, H.B. Weber {\it et al},
Appl. Phys. Lett. {\bf 82}, 4137 (2003)

%10
\bibitem{Sci03KDH}
S. Kubatkin, A. Danilov, M. Hjort {\it et al}, Nature {\bf 425}, 698 (2003)

%11
\bibitem{NL03KNY}
J.G. Kushmerick, J. Naciri, J. C. Yang, and R. Shashidhar,
Nano Lett. {\bf 3}, 897 (2003)

%12
\bibitem{PPC03PRL}
L. Patrone, S. Palacin, J. Charlier {\it et al}, Phys. Rev. Lett.
{\bf 91}, 096802 (2003)

%13
\bibitem{APL01SHI}
H. Sakaguchi, A. Hirai, F. Iwata {\it et al},
Appl. Phys. Lett. {\bf 79}, 3708 (2001)

%14
\bibitem{JPCB02CPZ}
X.D. Cui, A. Primak, X. Zarate {\it et al}, J. Phys. Chem. B
{\bf 106}, 8609 (2002)

%15
\bibitem{PRB03WLR}
W. Wang, T. Lee, and M.A. Reed, Phys. Rev. B {\bf 68}, 035416
(2003)

%16
\bibitem{Sci03XB}
B. Xu and N.J. Tao, Science {\bf 301}, 1221 (2003)

%17
\bibitem{JPCB02WHR}
D.J. Wold and C.D. Frisbie, J. Am. Chem. Soc. {\bf 123}, 5549 (2001);
D.J. Wold, R. Haag, M.A. Rampi {\it et al}, J. Phys. Chem. B
{\bf 106}, 2813 (2002)

%18
\bibitem{JPCB02AM} F. Anariba and R.L. McCreery, J. Phys. Chem. B
{\bf 106}, 10355 (2002)

%19
\bibitem{JPCB02IMA} T. Ishida, W. Mizutani, Y. Aya
{\it et al}, J. Phys. Chem. B {\bf 106}, 5886 (2002)

%THEORY
%21
\bibitem{PRB601LA}
N.D. Lang and Ph. Avouris, Phys. Rev. B {\bf 64}, 125323 (2001)

%22
\bibitem{JCP02NGI}
A. Nitzan, M. Galperin, G-L. Ingold, and H. Grabert, J. Chem. Phys. {\bf 117},
10837 (2002)

\bibitem{JCP02LWF}
Y. Luo, C-K. Wang, and Y. Fu, J. Chem. Phys. {\bf 117}, 10283 (2002)

%23
\bibitem{PRB03CLW}
H. Chen, J.Q. Lu, J. Wu {\it et al}, Phys. Rev. B {\bf 67}, 113408 (2003)

%24
\bibitem{PRB03XR}
Y. Xue and M.A. Ratner, Phys. Rev. B {\bf 68}, 115406 (2003); {\it ibid},
{\bf 68}, 115407 (2003)

%25
\bibitem{CMS03STB}
K. Stokbro, J. Taylor, M. Brandbyge {\it et al},
Comp. Mat. Sci. {\bf 28}, 151 (2003)


%26
\bibitem{JPC03MND}
V. Mujica, A. Nitzan, S. Datta {\it et al},
J. Phys. Chem. {\bf 107}, 91 (2003)


%27
\bibitem{PRB96STD}
M.P. Samanta, W. Tian, S. Datta {\it et al},
Phys. Rev. B {\bf 53}, R7626 (1996)

%28
\bibitem{CMS03NLG}
M. Nolan, J.A. Larsson, and J.C. Greer, Comp. Mat. Sci. {\bf 27}, 166 (2003)

%29
\bibitem{PRB03KLG}
C.C. Kaun, B. Larade, and H. Guo, Phys. Rev. B {\bf 67}, 121411 (2003);
C.C. Kaun and H. Guo, Nano Lett. {\bf 3}, 1521 (2003)

%30
\bibitem{JCP04TS}
J.K. Tomfohr and O.F. Sankey, J. Chem. Phys. {\bf 120}, 1542 (2004)


%31
\bibitem{PRL00VPL}
M. Di Ventra, S.T. Pantelides, and N.D. Lang,
Phys. Rev. Lett. {\bf 84}, 979 (2000)

%32
\bibitem{PRB01KB}
P.E. Kornilovitch and A.M. Bratkovsky, Phys. Rev. B {\bf 64}, 195413 (2001)

%33
\bibitem{PRB01FCR}
G. Fagas, G. Cuniberti, and K. Richter, Phys. Rev. B {\bf 63}, 045416 (2001)

%34
\bibitem{CP02CFR}
G. Cuniberti, G. Fagas, and K. Richter, Chem. Phys. {\bf 281}, 465 (2002)

%35
\bibitem{JACS02SDB}
J. M. Seminario, P. A. Derosa, and J. L. Bastos,
J. Am. Chem. Soc. {\bf 124}, 10266 (2002)


%36
\bibitem{EPL03GFR}
R. Gutierrez, G. Fagas, K. Richter {\it et al}, Europhys. Lett. {\bf 62},
90 (2003)

%37
\bibitem{PRL03EK}
E.G. Emberly and G. Kirczenow, Phys. Rev. Lett. {\bf 91}, 188301 (2003)

%38
\bibitem{PRL03HWW}
M.H. Hettler, W. Wenzel W, M.R. Wegewijs {\it et al},
Phys. Rev. Lett. {\bf 90}, 076805 (2003)

%39
\bibitem{JPCB03TY}
T. Tada and K. Yoshizawa, J. Phys. Chem. B {\bf 107} 8789 (2003)

%40
\bibitem{CPC03GGS}
R. Gutierrez, F. Grossmann, and R. Schmidt, ChemPhysChem {\bf 4},
1252 (2003)

%\bibitem{PRB02GFC} R. Gutierrez, G. Fagas, G. Cuniberti {it et al}, Phys. Rev. B {\bf 65}, 113410 (2002)


% THEORY OF TUNNELLING
%37
\bibitem{JCP94MKR}
V. Mujica, M. Kemp, and M.A. Ratner, J. Chem. Phys. {\bf 101} 6856 (1994) 

\bibitem{SSC98OKM}
A. Onipko, Y. Klymenko, L. Malysheva, and S. Stafstr{\"o}m,
Solid State Comm. {\bf 108}, 555 (1998)

\bibitem{PRB98MJ}
M. Magoga and C. Joachim, Phys. Rev. B {\bf 57}, 1820 (1998)

\bibitem{EPL96JV}
C. Joachim and J.F. Vinuesa, Europhys. Lett. {\bf 33}, 635 (1996)

\bibitem{PRB97MJ}
M. Magoga and C. Joachim, Phys. Rev. B {\bf 56}, 4722 (1997)

%Electron transfer
\bibitem{JCP61McC}
H.M. McConell, J. Chem. Phys. {\bf 35}, 508 (1961)

\bibitem{JCP92EK}
J.W. Evenson and M. Karplus, J. Chem. Phys. {\bf 96}, 5272 (1992)

\bibitem{JPPA94RH}
J.R. Reimers and N.S. Hush, J. Photochem. Photobiol. A {\bf 82}, 31 (1994)

\bibitem{JPC01N}
A. Nitzan, J. Phys. Chem. A {\bf 105}, 2677 (2001)

%RMT
\bibitem{CP98GPB}
E. Gudowska-Nowak, G. Papp, and  J. Brickmann,
Chem. Phys. {\bf 232}, 247 (1998)

\bibitem{CPL03LJ}
A. Lahmidi and C. Joachim, Chem. Phys. Lett. {\bf 381}, 381 (2003)

%effective mass and complex-band structure
\bibitem{CP02JM}
C. Joachim and M. Magoga, Chem. Phys. {\bf 281} 347 (2002)

\bibitem{PRB02TS}
J.K. Tomfohr and O.F. Sankey, Phys. Rev. B {\bf 65}, 245105 (2002)

\bibitem{PR59K}
W. Kohn, Phys. Rev. {\bf 115}, 809 (1959)

\bibitem{PPS63H}
V. Heine, Proc. Phys. Soc. {\bf 81}, 300 (1963)

\bibitem{Datta}
S. Datta, {\it Electronic Transport in Mesoscopic Systems},
Cambridge University Press, Cambridge (1995)

%3a
\bibitem{ARPC01N}
A. Nitzan, Annu. Rev. Phys. Chem. {\bf 52}, 681 (2001)

%resonances and band structure
\bibitem{CPL04FKE}
G. Fagas, A. Kambili, and M. Elstner, to be published in Chem. Phys. Lett.,
cond-mat/0308114

\bibitem{JPA99BG}
F. Barra and P. Gaspard, J. Phys. A: Math. Gen. {\bf 32}, 3357 (1999)

%GRAPHITE
\bibitem{CEJ02SBB}
C.D. Simpson, J.D. Brand, A.J. Berresheim {\it et al},
Chem.-Eur. J. {\bf 8}, 1424 (2002)

\bibitem{JPSJJ96FWN}
M. Fujita, K. Wakabayashi, K. Nakada, and K. Kusakabe,
J. Phys. Soc. Japan {\bf 65}, 1920 (1996)

\bibitem{PRB96NFD}
K. Nakada, M. Fujita, G. Dresselhaus, and M.S. Dresselhaus,
Phys. Rev. B {\bf 54},17954 (1996)

\bibitem{PRB99MNF}
Y. Miyamoto, K. Nakada, and M. Fujita, Phys. Rev. B
{\bf 59}, 9858 (1999) 

\bibitem{PRB00KMS}
T. Kawai, Y. Miyamoto, O. Sugino, and Y. Koga, Phys. Rev. B {\bf 62},
R16349 (2000)

\bibitem{JPSJ01SLC} F.L. Shyu, M.F. Lin, C.P. Chang {\it et al},
J. Phys. Soc. Japan {\bf 70}, 3348 (2001)

%'Unconventional' resonances
\bibitem{MS04FGR}
G. Fagas, R. Gutierrez, K. Richter {\it et al}, to be published
in Macromol. Symp.

%TB-DFT
\bibitem{TBDFT}
T. Frauenheim, G. Seifert, M. Elstner {\it et al},
Phys. Stat. Sol. (b) {\bf 217}, 41 (2000)

\bibitem{PRB98EPJ}
M. Elstner, D. Porezag, G. Jungnickel {\it et al},
Phys. Rev. B {\bf 58}, 7260 (1998)

\bibitem{SM03PMC}
A. Pecchia, M. Gheorghe, A. Di Carlo, and P. Lugli, Synth. Met. {\bf 138},
89 (2003)

\bibitem{JMS04EFS}
M. Elstner, T. Frauenheim, and S. Suhai,
to be published in J. Mol. Struct. (THEOCHEM)
 
%TRANSPORT
\bibitem{PRB99SLJ}
S. Sanvito, C.J. Lambert, J.H. Jefferson, and A.M. Bratkovsky,
Phys. Rev. B {\bf 59}, 11936 (1999)

%Electronic Structure
\bibitem{PRB81LJ} D.H. Lee and J.D. Joannopoulos, Phys. Rev. B {\bf 23}, 
4988 (1981)

\bibitem{PRB95AMV} C. Ambrosch-Draxl, J.A. Majewski, P. Vogl,
and G. Leising, Phys. Rev. B {\bf 51}, 9668 (1995)

\bibitem{JCPA01GGR}
S. Guha, W. Graupner, R. Resel {\it et al}, J. Phys. Chem. A {\bf 105},
6203 (2001)

\bibitem{PRL00LBS}
U. Landman, R.N. Barnett, A.G. Schebakov {\it et al},
Phys. Rev. Lett. 85 (2000) 1958

\bibitem{JCP03PSS}
S. Piccinin, A. Selloni, S. Scandolo {\it et al},
J. Chem. Phys. {\bf 119}, 6729 (2003)

\bibitem{JCM03PSD}
F. Picaud, A. Smogunov, A. Dal Corso {\it et al},
J. Phys.: Condens. Matter {\bf 15}, 3731 (2003)

\end{thebibliography}
\end{document}